\documentclass[twocolumn]{aastex631}
\shorttitle{Axial- and Equatorial-Dipole Coronal Holes}
\shortauthors{Panasenco, Wang, Badman et al.}

\begin{document}

\title{Axial-Dipole and Equatorial-Dipole Coronal Holes:\\ Congruent Magnetic Geometry, Divergent Evolution}

\correspondingauthor{Olga A. Panasenco}
\email{opanasenco@setiap.org}

\author[0000-0002-4440-7166]{Olga A. Panasenco}
\affiliation{Advanced Heliophysics Inc., Pasadena, CA, USA}

\author[0000-0002-3527-5958]{Yi-Ming Wang}
\affiliation{US Naval Research Laboratory, Washington, DC 20375, USA}

\author[0000-0002-6145-436X]{Samuel T. Badman}
\affiliation{Center for Astrophysics, Harvard \& Smithsonian, Cambridge, Massachusetts, 02138, USA}

\author[0000-0002-3020-9409]{Adam J. Finley}
\affiliation{European Space Agency, ESTEC, Noordwijk, The Netherlands}

\author[0000-0002-8748-2123]{Yeimy J. Rivera}
\affiliation{Center for Astrophysics, Harvard \& Smithsonian, Cambridge, Massachusetts, 02138, USA}

\begin{abstract}

While the Sun's polar fields reverse, the axial dipole is weak and
low-latitude coronal holes are common. Equatorial-dipole coronal holes (EDCHs)
appear when the equatorial dipole is strong, forming at its
magnetic poles, similar to the axial-dipole coronal holes (ADCHs) of minimum.
Not every low-latitude hole qualifies: the 2024 maximum was quadrupole-led.
Parker Solar Probe found fast wind from an EDCH deeply sub-Alfv\'enic at
$9.9\,R_\odot$, at only $60\%$ of an asymptotic speed matching polar wind,
which is usually thought to accelerate faster. We ask whether coronal structure
predicts this, using potential-field models of synthetic and observed fields
at three cycle phases. We show that pure axial and equatorial dipoles are
exactly congruent, with the same open areas and core expansion factors, so any
difference must arise from orientation relative to the differential-rotation
pattern or from the surrounding multipoles. We find that ADCHs are stationary,
sustained by converging meridional flow, whereas EDCHs depend on an equatorial
dipole that active-region emergence builds and flux transport dismantles: over
$18$-day sequences the axial-dipole strength at minimum is steady to $1\%$,
while the equatorial dipole varies by $21\%$ at maximum and $8\%$ in the
declining phase. Most open flux in EDCHs expands less than in ADCHs, but their
boundaries are pseudostreamer-dominated, with non-monotonic expansion. Despite
these differences, magnetic geometry alone does not predict slower acceleration in EDCHs; we
argue that any real difference points to extended heating by
reflection-driven Alfv\'en-wave turbulence, perhaps fed differently at those
boundaries, which potential-field models cannot capture.

\end{abstract}


\section{Introduction} \label{sec:intro}

Coronal holes are conventionally classified by latitude and area
\citep{Cranmer2009,Wang2009}. Two particular configurations share the same nominal magnetic geometry, while differing in nearly every property that governs their evolution. The first are the polar coronal holes at solar minimum, the footprints of the axial dipole; we call them \emph{axial-dipole coronal holes} (ADCHs). The second configuration appears during the solar cycle polarity reversal, when the axial dipole nulls and the equatorial dipole, anchored in a small number of large active-region (AR) complexes, becomes the dominant low-order component \citep{WangSheeley2003,DeRosa2012,Wang2014,Sun2015}. The dipole axis then lies in or near the equatorial plane. The solar wind emerging from such holes has been described as polar-hole-like \citep{Badman2026}, and the resemblance is real: the outflow is asymptotically indistinguishable from ADCH fast wind. We
nonetheless avoid framing these holes as merely resembling polar or ADCHs. As Section~\ref{sec:congruence} shows, where the equatorial dipole leads the low-order field these holes are not merely \emph{like} polar holes (ADCHs) but congruent with the poles of the non-axisymmetric dipole , and a description in terms of resemblance obscures
the point. We call them \emph{equatorial-dipole coronal
holes} (EDCHs), a name that identifies the harmonic component responsible and
therefore the observable that constrains their lifetime.

\begin{figure*}
\centering
\includegraphics[width=\textwidth]{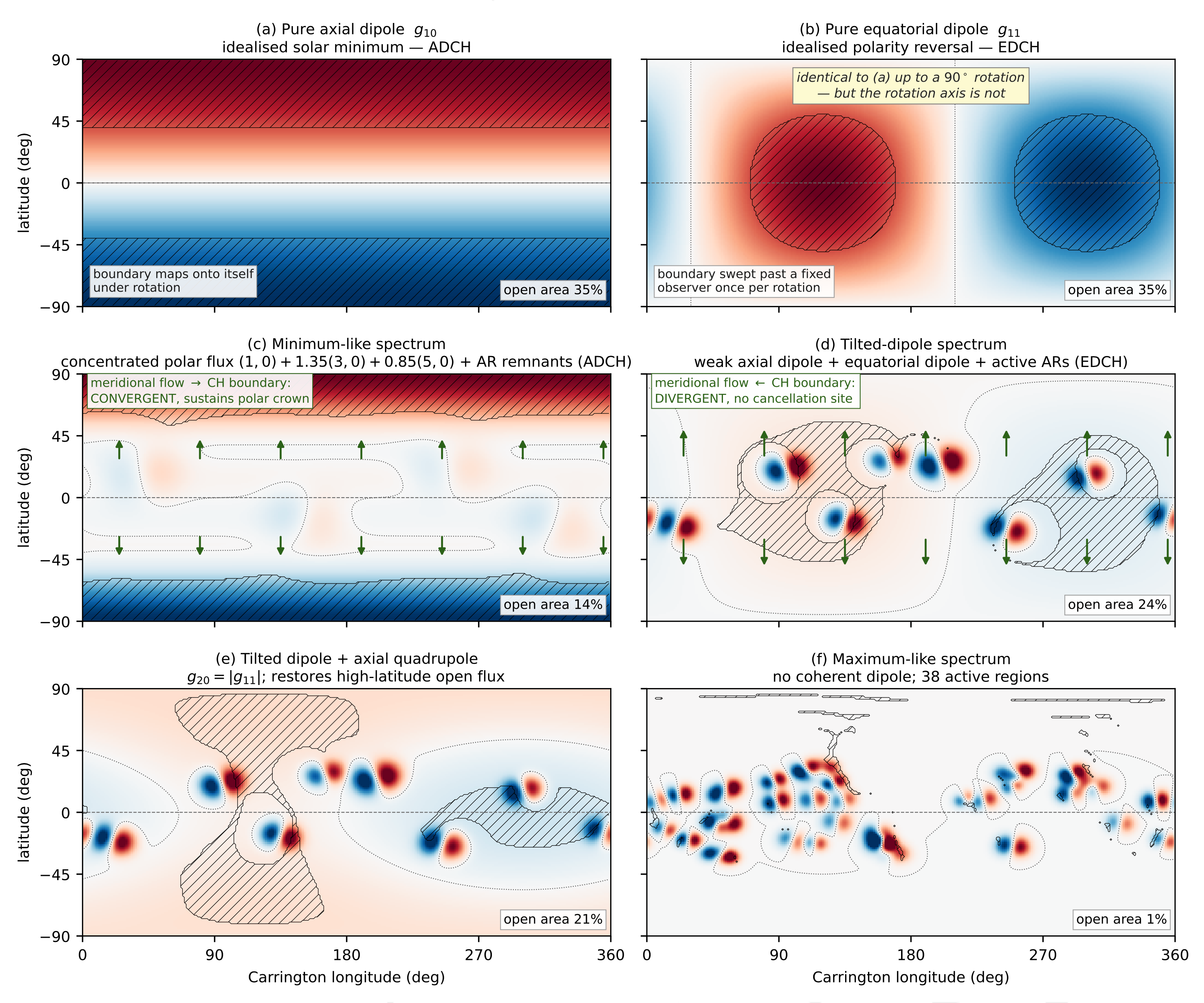}
\caption{Synthetic PFSS open-field maps for the six configurations described
in Section~\ref{sec:models}, computed with a source surface at
$R_{ss}=2.5\,R_\odot$. Color: photospheric $B_r$, red positive, each panel
scaled independently. Hatching and outline: open-field region. Dotted:
photospheric PILs. Green arrows: sense of the surface meridional flow.
Panels (a) and (b) are the same field rotated by $90^\circ$ and yield identical
open areas and expansion factors. In panel (c) the polar flux is concentrated in the ratio
$(1,0)+1.35\,(3,0)+0.85\,(5,0)$ of Schmidt-normalized coefficients measured from the 2019 October maps, giving an
open area of $14\%$ and a boundary near $53^\circ$; a pure dipole would place
the boundary at $40^\circ$ and open $35\%$ of the surface. Panels (a) and (c)
correspond to axial-dipole coronal holes, panel (b) to an equatorial-dipole
hole, and panels (d) and (e) to the tilted-dipole configurations in which such
holes are found. Panel (e) adds an axial quadrupole of amplitude equal to the
equatorial dipole, restoring high-latitude open flux without restoring the
axial dipole. Panel (f) suppresses the coherent low-order field entirely and
the open flux collapses into many disconnected fragments, the regime
characteristic of solar maximum.\label{fig:maps}}
\end{figure*}

During the global polarity reversal the solar magnetic axis does not sweep smoothly through the equator \citep[See Fig. 1 in ][]{Finley2024}. The equatorial dipole is a distinct component whose strength and orientation change as its source complexes emerge and decay, a fact we return to in Section~\ref{sec:conclusions} as one of the timescales limiting the lifetime of an EDCH. Differential rotation shears the equatorial dipole into higher-order multipoles, which supergranular diffusion and poleward meridional flow then annihilate through flux cancellation on a $\sim1$~yr timescale. The equatorial dipole must therefore be continually regenerated by new active regions, with a strength that depends on the longitudinal and hemispherical distribution of flux emergence \citep{Wang2014}. 

We argue that ADCHs and EDCHs are not species of the same object at different latitudes. This highlights the central difference between ADCHs and EDCHs: the axial-dipole hole is a \emph{stationary state} of the global system, whereas the equatorial-dipole hole is a \emph{transient configuration} that inherits the same footpoint topology. Establishing that distinction, and determining which observed differences can and cannot be attributed to it, is the first aim of this paper.
 
EDCHs produce wind asymptotically indistinguishable from polar fast wind, with speeds above $700\,\mathrm{km\,s^{-1}}$ by
$\sim60\,R_\odot$, $\sim750\,\mathrm{km\,s^{-1}}$ at 1~au, and coronal proton and electron temperatures consistent with SOHO/UVCS determinations in polar holes
\citep{Kohl1998,Cranmer2020}. Yet the measured acceleration profile is markedly extended, only $\sim60\%$ complete by $10\,R_\odot$ \citep{Badman2026, Dakeyo2026},
in apparent conflict with interplanetary scintillation (IPS) results indicating that the fast polar wind reaches its final velocity inside $10$--$15\,R_\odot$ \citep{Grall1996,Breen1996}. Both studies conditioned that conclusion on the scintillating density irregularities moving at the bulk flow speed, an assumption their authors flagged and left open. Other near-Sun IPS measurements, moreover, find the wind still accelerating between $10$ and $30\,R_\odot$ \citep{Tokumaru1994,Tokumaru1996}, and UVCS measurements have shown outflow in an equatorial coronal hole three times slower than in polar holes at $2$--$3\,R_\odot$, despite similar asymptotic speeds \citep{Miralles2001}. Resolving that tension is the second aim of this paper.

\section{Synthetic PFSS Models} \label{sec:models}

\begin{figure*}
\centering
\includegraphics[width=\textwidth]{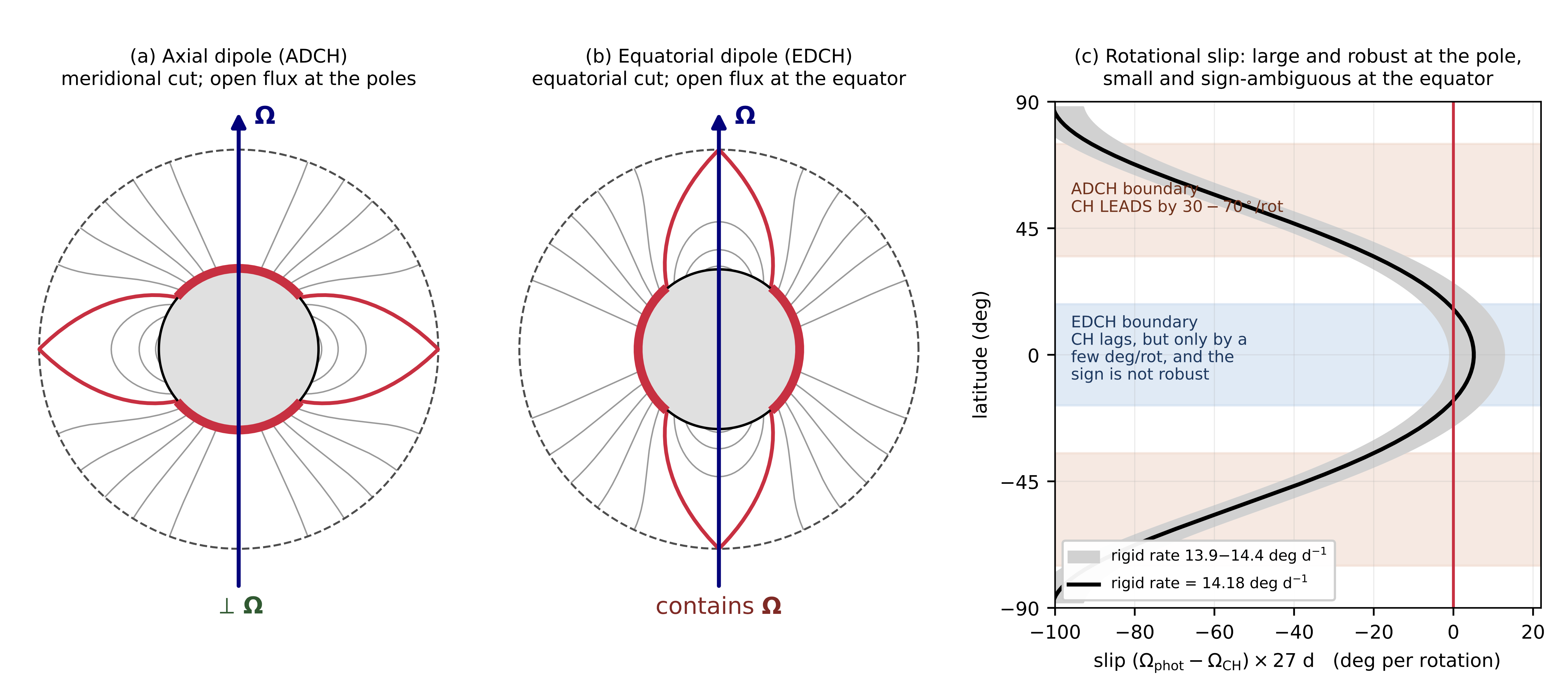}
\caption{Congruent magnetic geometry, opposite relation to the rotation axis.
(a, b) Dipole field lines in the plane containing the dipole axis, with the
rotation axis $\mathbf{\Omega}$ marked; crimson denotes the separatrix and the
open footpoint arcs. Panel (b) is the same field as panel (a) rotated by
$90^\circ$, so the two are magnetically congruent, yet they stand in opposite
relation to $\mathbf{\Omega}$. The axial-dipole hole boundary is perpendicular
to $\mathbf{\Omega}$, so rotation maps it onto itself and every boundary
process can relax to a quasi-steady state. The equatorial-dipole hole boundary
contains $\mathbf{\Omega}$, so it crosses latitudes: differential rotation
slips the photospheric plasma across it at its poleward flanks, interchange
reconnection must continually hold it in place, and no steady state is
reached. (c) Rotational
slip $(\Omega_{\rm phot}-\Omega_{\rm CH})$ per rotation, using the
\citet{Snodgrass1983} magnetic rate; the grey band spans plausible quasi-rigid
CH rates. \label{fig:geom}}
\end{figure*}

We construct idealized photospheric $B_r$ distributions and solve the
potential-field source-surface problem with $R_{ss}=2.5\,R_\odot$, expanding in
orthonormal real spherical harmonics and obtaining $\mathbf{B}=-\nabla\Phi$.
The photospheric field is represented on a $145\times288$
latitude--longitude grid ($1.24^\circ$ in colatitude, $1.25^\circ$ in
longitude), with harmonics retained to $\ell_{\max}=6$ for the pure-dipole
cases and $\ell_{\max}=22$ for the realistic spectra of configurations
(c)--(f) and the sequence f1--f5 of Table~\ref{tab:metrics}. For the
open-field metrics of Table~\ref{tab:metrics}, whose component counts are
sensitive to truncation, the expansion is carried to $\ell_{\max}=30$.
The scalar potential
is evaluated on a three-dimensional grid spanning $1$--$2.5\,R_\odot$ in $64$
radial steps, $\mathbf{B}$ is obtained from it by second-order centred
differences, and the Cartesian components are interpolated trilinearly for
tracing. Open-field footpoints are identified by tracing a field line from
each of the $41{,}760$ grid nodes with a fourth-order Runge--Kutta integrator
at a fixed arclength step of $0.02\,R_\odot$ and testing whether it reaches
$R_{ss}$. For the open-area maps the seeds are uniform in latitude and longitude, so high
latitudes are oversampled and all areal statistics are $\sin\theta$-weighted.
For the expansion-factor analysis of Section~\ref{sec:expansion} the seeds are
instead distributed uniformly in solid angle, with rows spaced uniformly in
$\cos\theta$, so that every seed carries the same area; on a uniform
latitude--longitude grid a polar cap collects far more seeds per unit area than
an equatorial hole of the same size, which biases any comparison between them.
That tracing also refines the radial sampling to $200$ steps and reduces the
integration step to $0.006\,R_\odot$, to resolve expansion maxima below the
source surface. The calculations use NumPy \citep{Harris2020} and SciPy
\citep{Virtanen2020}, and Figures~\ref{fig:maps}--\ref{fig:ips} were
produced with Matplotlib \citep{Hunter2007}. The models are idealized and are
not intended to reproduce any particular Carrington rotation. A spherical source surface at a fixed radius
is itself an idealization
\citep{Panasenco2020}; we return to its consequences in
Section~\ref{sec:expansion}.

We denote a spherical-harmonic component of degree $l$ and order $m$ by $(l,m)$
and its coefficient by $g_{lm}$; coefficients are Schmidt semi-normalized and
quoted in gauss throughout (Appendix~\ref{app:harmonics}), so that $g_{10}$ is
the polar field strength of the equivalent axial dipole. Thus $(1,0)$ is the
axial dipole, $(1,1)$ the equatorial dipole and $(2,2)$ a sectoral quadrupole. Six configurations are
considered and shown in Figure~\ref{fig:maps}: (a) a pure axial dipole; (b) a pure equatorial dipole; (c) a minimum-like spectrum in which the polar flux is concentrated as $(1,0)+1.35\,(3,0)+0.85\,(5,0)$, plus decayed low-latitude bipoles; (d) a tilted-dipole spectrum with a weak axial dipole, an equatorial dipole, and eight active magnetic bipoles at low latitude; (e) the tilted-dipole spectrum with an added axial quadrupole of amplitude equal to the equatorial dipole, motivated by the observed polar fields of Section~\ref{sec:obsharm}; and (f) a maximum-like spectrum with no coherent dipole and 38 active regions, the most fragmented member (f5) of the sequence f1--f5 of Table~\ref{tab:metrics}. The structural analysis of Sections~\ref{sec:structure}--\ref{sec:wind} rests on (a)--(d); configurations (e) and (f) enter through Table~\ref{tab:metrics} and Section~\ref{sec:obsharm}.

The polar profile of configuration (c) requires comment, because a pure dipole
is a poor description of solar minimum. Meridional transport concentrates the
polar flux, so the axisymmetric field is not $(1,0)$ alone and the polar holes
reach only to about $60^\circ$ instead of $40^\circ$. We take the concentration
from measurement rather than assumption. Decomposing five of our maps spanning
2019 October 1--19, near the cycle 24/25 minimum, gives
$g_{30}/g_{10}=1.35$ and $g_{50}/g_{10}=0.70$, stable to a few per cent across
the five, and configuration (c) adopts that ratio. It has an open-area fraction
of $0.14$ and a polar hole reaching to $53\fdg2$ once the active-region
remnants are included.

Two checks support that profile. The profile corresponds to a polar field
$B_r\propto\cos^{n}\theta$ with $n=7.9$, within the range $n\sim7$--$9$
conventionally assumed for polar caps \citep{Svalgaard1978}, and reproduces the
second ratio, $g_{50}/g_{10}=0.74$ against the measured $0.70$, which was not
used to fix $n$. And it is consistent with the low-order source-surface
components measured from Wilcox Solar Observatory (WSO) data
\citep{Hoeksema1995} across four minima \citep{Wang2014}, which
give $g_{30}/g_{10}$ in the range $1.4$--$2.2$ once scaled to the photosphere.
Our value sits just below the low end of that range, in the direction expected if the
WSO polar treatment concentrates the flux somewhat more than the polar filling
used here.  The pure dipoles of panels (a) and (b) of Figure~\ref{fig:maps} are retained only for the congruence argument of Section~\ref{sec:congruence}, where the point is an exact identity between two idealized fields that any added multipole would obscure; they are not intended as representations of solar minimum.

The solver reproduces the analytic axial-dipole result, for which the open
boundary lies at latitude
$\lambda_b = 90^\circ-\arcsin\!\left\{\left[3\tilde{R}_{ss}^{2}/\bigl(1+2\tilde{R}_{ss}^{3}\bigr)\right]^{1/2}\right\} = 40.3^\circ$,
with $\tilde{R}_{ss}\equiv R_{ss}/R_\odot$, and the open-area fraction is
$0.353$. The numerical solution gives an open-area fraction of $0.354$,
equivalent to a boundary latitude of $40.2^\circ$. The innermost grid row
containing open flux lies at $40.9^\circ$, half a cell poleward of the
analytic boundary and within the one-cell ($1.24^\circ$) offset expected
when a smooth boundary is resolved on a discrete grid. We quote the
area-equivalent latitude, because the area is integrated directly whereas
the row position is quantised at the grid spacing.

\subsection{Congruence of the Pure Dipoles} \label{sec:congruence}

Panels (a) and (b) of Figure~\ref{fig:maps} give open-area fractions of
$35.4\%$ and $35.3\%$, and the core-tube expansion factors computed in
Section~\ref{sec:expansion} are $1.724$ and $1.720$, against an analytic value
of $1.720$ for a pure dipole. The residual differences, $0.1\%$ in area and
$0.2\%$ in the core expansion factor, are numerical rather than physical: the
pole of the axial dipole coincides with a coordinate pole, near which
equal-area seeding places few rows, so its innermost seed lies $3.3^\circ$ from
the pole, whereas the pole of the equatorial dipole lies in the interior of the
grid. The agreement is an identity: the two fields are related by a rigid rotation, and the PFSS operator is rotationally invariant. \emph{For a pure dipole there is no magnetic distinction whatever between an axial-dipole and an equatorial-dipole coronal hole.}

This sharpens the problem considerably. Every difference between the two CH classes must originate in one of exactly two places: the orientation of the CH boundary relative to the rotation axis $\mathbf{\Omega}$, and hence to the differential-rotation pattern, which the PFSS operator does not see; and the multipolar environment in which the dipole sits, which distinguishes panels (c) and (d). The two turn out to govern different things. Orientation governs how long a hole persists (Section~\ref{sec:structure}); the environment governs its structure, and in particular gives the equatorial-dipole hole the pseudostreamer-dominated boundary that is the one structural feature separating the classes (Section~\ref{sec:expansion}).

\begin{deluxetable*}{lccc}
\tablecaption{Stability of the low-order photospheric field over
$16$--$18$~day intervals at three phases of the solar cycle. Each entry is the
mean and fractional standard deviation of five maps at $4$--$6$ day spacing,
decomposed identically on the native Gauss--Legendre grid
(Appendix~\ref{app:harmonics}). Coefficients are Schmidt semi-normalized and in
gauss, so that $g_{10}$ is the polar field strength of the equivalent axial
dipole. The phase drift is that of the equatorial dipole over the
interval.\label{tab:stability}}
\tablehead{\colhead{} & \colhead{2019 Oct 1--19} & \colhead{2024 Sep 1--19} &
\colhead{2025 Mar 12--28} \\
\colhead{} & \colhead{CR 2222--2223} & \colhead{CR 2288--2289} &
\colhead{CR 2295} \\
\colhead{} & \colhead{(minimum, cycles} & \colhead{(maximum,} &
\colhead{(declining,} \\
\colhead{} & \colhead{24--25)} & \colhead{cycle 25)} & \colhead{cycle 25)}}
\startdata
dominant low-order component & $g_{10}$ & $|g_{11}|$ & $|g_{11}|$ \\
its mean amplitude (G)       & $1.20$   & $1.62$     & $2.03$ \\
its fractional s.d.          & $1\%$    & $21\%$     & $8\%$ \\
\hline
$g_{10}$ (G)                 & $+1.20$, $1\%$     & $+0.59$, $32\%$    & $+0.39$, $48\%$ \\
$|g_{11}|$ (G)               & $\phn0.12$, $16\%$ & $\phn1.62$, $21\%$ & $\phn2.03$, $\phn8\%$ \\
$|g_{22}|$ (G)               & $\phn0.26$, $25\%$ & $\phn1.51$, $35\%$ & $\phn2.86$, $13\%$ \\
$g_{30}$ (G)                 & $+1.62$, $1\%$     & \nodata           & \nodata \\
$g_{50}$ (G)                 & $+0.85$, $6\%$     & \nodata           & \nodata \\
mean unsigned $B_r$ (G)      & $\phn2.29$, $4\%$  & $\phn9.88$, $7\%$  & $\phn7.48$, $\phn3\%$ \\
power in $l=1$               & $49$--$53\%$       & $9$--$19\%$        & $20$--$34\%$ \\
equatorial dipole phase drift & \nodata           & $31^\circ$         & $7^\circ$ \\
\enddata
\tablecomments{The 2019 interval precedes the cycle 24/25 minimum, usually
dated to 2019 December, by about two months. Polar fields are steady across it,
varying by less than $1\%$ over the five maps ($+1.84$ to $+1.78$~G north,
$-3.18$ to $-3.21$~G south). The large fractional scatter in $g_{10}$ in
2025 March reflects its small amplitude rather than rapid change: the absolute
scatter, $0.19$~G, is smaller than in 2024 September. The residual scatter in
$|g_{22}|$ persists at all three phases and we base no argument on that
coefficient. The axisymmetric terms $g_{30}$ and $g_{50}$ are quoted only for
the minimum interval, where they describe the polar cap and give
$g_{30}/g_{10}=1.35$ and $g_{50}/g_{10}=0.71$, the ratios adopted for
configuration (c) in Section~\ref{sec:models}. At the other two phases the same
coefficients are dominated by active-region flux at low latitudes rather than by
polar structure, and are not interpretable as a description of the polar
field.}
\end{deluxetable*}

\section{Structural Distinctions} \label{sec:structure}

\subsection{Rotational Symmetry and Differential-Rotation Shear}
\label{sec:symmetry}

An axial-dipole hole is invariant under the rotation the system is actually performing: rotation maps its boundary onto itself (Figure~\ref{fig:geom}a). Every boundary process (tangential shear, interchange reconnection, flux delivery from below) can therefore relax to a quasi-steady state. The observed quasi-rigid rotation of polar CHs and of their equatorward extensions \citep{Timothy1975,Wang1988,WangSheeley1993} goes beyond what this symmetry argument alone predicts, since the extensions are non-axisymmetric; we invoke those observations here only as evidence that ADCH boundaries do reach a statistically steady configuration.

The equatorial dipole has its axis \emph{in} the rotation plane (Figure~\ref{fig:geom}b), and its anchoring active-region complexes rotate with the photosphere, so the rotation of the pattern as a whole has no effect by itself. Nor is the hole progressively deformed. A coronal hole rotates at the rate of the global nonaxisymmetric photospheric flux \citep{WangSheeley1993}: when that flux lies at low latitudes, as it commonly does on the declining phase, even the midlatitude parts of the hole rotate at the equatorial rate. Coronal holes are the footpoints of field lines extending far into the heliosphere, whose sources are the lowest-order multipoles of the photospheric field, and these rotate far more rigidly than the high-order multipoles. The contrast is large and measurable. Across the sequences of Table~\ref{tab:stability}, the effective multipole order $l_{\rm eff}=[\sum_l l(l+1)P_l/\sum_l P_l]^{1/2}$, with $P_l$ the power in degree $l\le20$, is $7.6$ to $14.1$ for the photospheric field but only $1.6$ to $2.2$ at the source surface, lowest at minimum and highest at maximum. Shearing converts low-order multipoles into high-order ones, and interchange reconnection along the boundary continually opposes it.

What differs between the two classes is therefore not whether the hole keeps its shape, but how its boundary lies relative to the differential-rotation pattern. The ADCH boundary runs at constant latitude, so the photosphere slips along it and reconnection has nothing to undo. The EDCH boundary crosses latitudes, so the photospheric plasma slips across it at the poleward flanks, by tens of degrees per rotation for a hole whose angular radius approaches $50^\circ$ (Section~\ref{sec:slip}), and interchange reconnection must operate continuously to hold the boundary in place. The same differential rotation acts on the harmonic component itself, winding the equatorial dipole into higher-order multipoles that supergranular diffusion and poleward meridional flow then annihilate through flux cancellation, which drives the $\sim1$~yr decay and continual regeneration documented by \citet{Wang2014}. Flux transport therefore acts on the equatorial-dipole hole only indirectly, through the harmonic component that sustains it. Together with the intrinsic evolution of the anchoring complexes (Section~\ref{sec:residence}), this keeps the EDCH configuration from reaching a steady state, even though at any moment its boundary is held quasi-rigid by reconnection.

\subsection{Flux Residence Time Rather than a ``Flux Sink''}
\label{sec:residence}

The polar cap is the terminus of the surface flux-transport circulation.
Poleward meridional advection delivers flux there and there is nowhere further for it to go; removal proceeds only by cancellation against arriving opposite polarity \citep{Babcock1961,Leighton1964,Wang1989,Sheeley2005,Jiang2014}. The
residence time is of order years.

Low latitudes are the \emph{source} region of the same circulation. Flux there has a residence time set by the meridional crossing time, roughly one year for $30^\circ$ at $10$--$15\,\mathrm{m\,s^{-1}}$
\citep{HathawayRightmire2010}, and shorter once supergranular diffusive
shredding among AR remnants is included. The equatorial-dipole coronal hole is rooted in flux that surface transport is actively dispersing. The lifetimes of the two CH classes are therefore set by different physical quantities.

This contrast is directly measurable in the low-order field, and the two
regimes can be compared on equal terms by decomposing sequences of maps at the
same cadence and spacing at each phase (Table~\ref{tab:stability}). Five maps
spanning 2024 September 1--19, near the sunspot maximum, give an equatorial
dipole $|g_{11}|$ with a standard deviation of $21\%$ about a mean of
$1.62$~G, while the mean unsigned photospheric field
falls monotonically from $10.6$ to $9.0$~G as the anchoring active regions
decay. Five maps spanning 2019 October 1--19, near the minimum between cycles 24 and 25,
give an axial dipole $g_{10}$ of $1.20$~G with a standard deviation of $1\%$,
the polar fields themselves varying by less than that. 

The dominant components have comparable amplitudes at the two phases, $1.20$
against $1.62$~G, but their amplitudes change on timescales differing by a factor
of twenty. That is what the flux-transport argument above predicts: one is fed
by a circulation that terminates at the pole and can be removed only by
cancellation, the other by active regions that emerge and decay in place. Behind that lies a difference in how active regions contribute
to the two components. Their axial-dipole moments accumulate, all pointing the
same way by Joy's law, so no single emergence changes the axial dipole much;
their contributions to the equatorial dipole are vectors in the equatorial
plane, whose sum depends on the longitudinal and hemispheric distribution of
the regions and so changes with each emergence \citep{WangSheeley2003}. It
also shows that the scatter at maximum is not an artifact of the map product,
since the same maps analysed the same way are stable to $1\%$ at minimum.

A third sequence, five maps spanning 2025 March 12--28 on the declining phase,
falls between the two. There $|g_{11}|$ has a mean of $2.03$~G and a standard
deviation of $8\%$, and its phase drifts by $7^\circ$ over sixteen days
against $31^\circ$ over eighteen days in September. The equatorial dipole is
by then both stronger and steadier than it was at maximum, though still an
order of magnitude less stable than the axial dipole at minimum. This is the
persistence that \citet{Finley2026} and \citet{Badman2026} describe for the
2025 tilted dipole, expressed as a measurement: the configuration settles once
the nests that anchor it are established, and it is the interval during which
it is being assembled, not the interval during which it exists, that is
volatile. The three phases therefore order as the taxonomy requires, with an
established equatorial-dipole configuration intermediate between a polar one
and one under construction.

The orientation is more stable than the amplitude. Over the same interval the
phase of the equatorial dipole drifts by only $31^\circ$, about
$1\fdg7\,\mathrm{day^{-1}}$ and far slower than rotation, so the component
retains its longitude while its strength varies with the emergence and decay of
the anchoring complexes. This is the continual-regeneration behavior described
by \citet{Wang2014}, and it refines the coherence time invoked in
Section~\ref{sec:conclusions}: the orientation of an equatorial-dipole
configuration persists for several rotations, but its strength does not.

The pattern is what anchoring in active-region nests would produce. Nests are
by construction persistent in Carrington longitude while their flux emerges
episodically \citep{FinleyEtal2025,Finley2026}, so a dipole rooted in them should
hold its orientation while its amplitude steps with each emergence. During 2024
three nests in the southern hemisphere accounted for most of the Sun's flaring
activity, and decaying nest flux is itself a recognized source of low-latitude
coronal holes \citep{MordvinovYazev2014,Finley2026}. The equatorial-dipole
hole is therefore anchored not merely in decaying active-region flux but in
flux that is repeatedly replenished at a fixed longitude, which is why its
orientation outlives any individual active region while its strength does
not.

\subsection{Boundary Flows and the Filament-Channel Asymmetry}
\label{sec:filaments}

At the polar crown polarity inversion line (PIL) the classical filament-channel formation recipe is satisfied continuously: a PIL located just equatorward of the polar CH boundary, sustained converging
flow into it as trailing-polarity flux is advected against the polar cap,
ongoing cancellation, and latitudinal shear acting \emph{along} an east--west channel
\citep{vanBallegooijenMartens1989,Martin1998,Martens2001,vanBallegooijen1998}.
Months of uninterrupted axial-field and helicity injection follow, which is why polar crown filaments obey the hemispheric chirality rule so strictly: the chirality is imposed by the global driver rather than inherited from a source AR \citep{Martin1994,Pevtsov2003,MackayvanBallegooijen2005,Panasenco2014}.
Their persistence over many rotations, poleward migration, and disappearance at reversal follow the same clock as the ADCH itself
\citep{Leroy1983,HarveyRecely2002,Mackay2010}.

At an equatorial-dipole hole the poleward boundary is instead a flow \emph{divergence} line (Figure~\ref{fig:maps}d): meridional advection carries flux across it and away, so the boundary migrates rather than pinning a PIL, and no sustained cancellation site is established. Channels that do form nearby inherit their chirality from the parent AR and possess no persistent driver to rebuild them after eruption.

\subsection{Topological Fragility} \label{sec:topology}

At solar minimum the axial-dipole hole boundary is essentially the single global separatrix, the photospheric trace of the heliospheric current sheet: one structurally stable surface. During reversal the equatorial-dipole hole boundary is composed largely of S-web quasi-separatrix layers and pseudostreamer separatrices, with much of
the open flux threaded through the pseudostreamer belt
\citep{Wang2007,Antiochos2011,Titov2011,Higginson2017}. Such a boundary is intrinsically leaky to interchange reconnection
\citep{Fisk2005,FiskKasper2020,Bale2023}, consistent with these regions
producing variable outflow with a substantial Alfv\'enic slow-wind component \citep{DAmicisBruno2015,DAmicis2019,DAmicis2021a,DAmicis2021b,DAmicis2025,Panasenco2020}. Slow, highly structured
wind has indeed been observed emerging from an equatorial coronal hole
\citep{Bale2019}.

\begin{figure*}
\centering
\includegraphics[width=\textwidth]{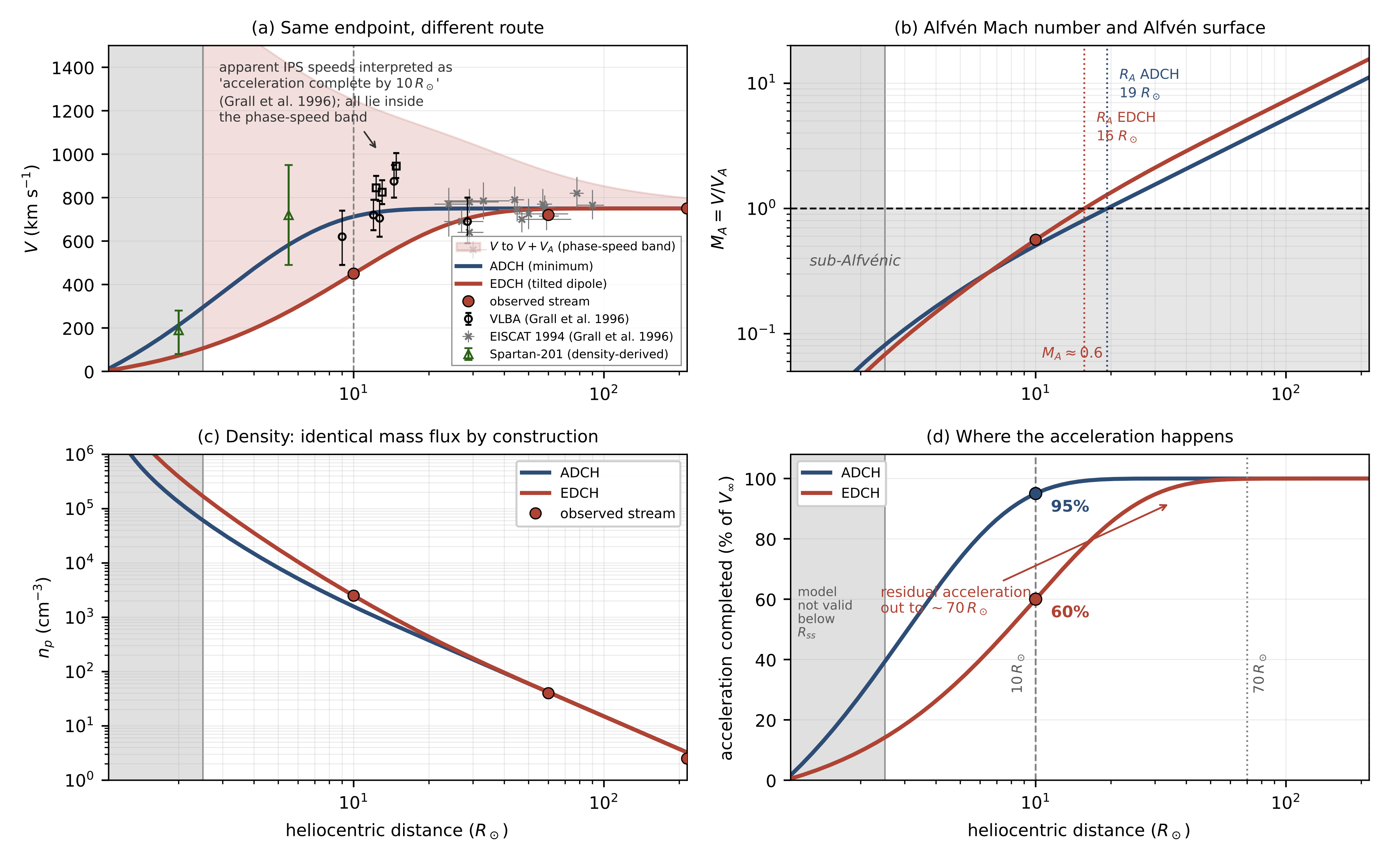}
\caption{Synthetic wind profiles for the two source classes at identical mass
flux, with $V_\infty=750\,\mathrm{km\,s^{-1}}$. Grey shading marks
$r<R_{ss}=2.5\,R_\odot$, where the $r^{-2}$ field scaling does not apply and the
curves should not be read. (a) Speed. The shaded band
spans $V$ to $V+V_A$ for the equatorial-dipole hole, the range within which a
scintillation tracer propagating at a fluctuation phase speed would fall.
Overlaid are the apparent speeds digitized from Figure~3 of \citet{Grall1996}:
VLBA measurements of the south and north polar streams (open circles and
squares), the 1994 EISCAT series (grey crosses, also shown by
\citealt{Breen1996}), and the density-derived Spartan-201 estimates at 2 and
$5.5\,R_\odot$ (green triangles). The scintillation points are digitized from
the published figure rather than taken from a table; the digitization
uncertainty is comparable to or smaller than the plotted error bars. Every
point lies inside the band. Filled circles are the measured stream. (b) Alfv\'en Mach number, with the
sub-Alfv\'enic region shaded and the Alfv\'en radius marked for each case.
(c) Proton density. (d) Fraction of the acceleration completed as a function
of radius. The profiles use the illustrative parameterisation
$V(r)=V_\infty\left[1-e^{-(r-1)/H}\right]$ with $H=3.0\,R_\odot$ for the
axial-dipole hole and $9.8\,R_\odot$ for the equatorial-dipole hole, chosen to
reproduce the measured stream, together with mass-flux conservation and
$B\propto r^{-2}$. The rapid axial-dipole profile represents the literal reading of the
scintillation results; Section~\ref{sec:ips} and Figure~\ref{fig:ips} compare it
with other near-Sun constraints. The axial-dipole field is taken to be stronger by a factor
$1.4$, which is an assumption rather than a result: since
$R_A^2V(R_A)=\Phi_B^2/\mu_0\Phi_m$, the Alfv\'en radius is fixed by the ratio
of magnetic to mass flux, and the larger $R_A$ shown for the axial-dipole case
follows directly from that choice. This is a parameterisation, not a solution
of the momentum equation.
\label{fig:wind}}
\end{figure*}

\subsection{Rotational Slip} \label{sec:slip}

Both CH classes rotate quasi-rigidly while the photosphere beneath rotates differentially \citep{Snodgrass1983,Nash1988}, so at high latitude the CH leads the photosphere and near the equator it lags. Quasirigid rotation itself requires that the coronal field be continually decoupled from the photospheric plasma by reconnection \citep{Wang2014}, consistent with the boundary picture of Section~\ref{sec:symmetry}. We report this asymmetry with a caveat (Figure~\ref{fig:geom}c). We use the \citet{Snodgrass1983} magnetic-feature rate,
$\Omega_{\rm phot}=14.37-2.30\sin^2\lambda-1.62\sin^4\lambda\ \mathrm{deg\,d^{-1}}$
(sidereal), converted from the microradian form
$2.902-0.464\sin^2\lambda-0.328\sin^4\lambda\ \mathrm{\mu rad\,s^{-1}}$ in which
\citet{Snodgrass1983} publishes it;
the slip is a rate difference, so the sidereal--synodic distinction cancels, and we quote it per $27.28$-d synodic rotation. The polar slip is large and robust: $37^\circ$ and $67^\circ$ per rotation at $45^\circ$ and $60^\circ$ latitude for the plotted rate, and $30$--$73^\circ$ across the plausible range of rigid rates. The equatorial slip is small but not negligible: $+5^\circ$ per rotation at the equator for the plotted rate, falling below $3^\circ$ in magnitude between $10^\circ$ and $20^\circ$ latitude, though it reaches $+13^\circ$ at the equator at the slow edge of the band. Its \emph{sign} is not robust: for the plotted rate of $14.18\,\mathrm{deg\,d^{-1}}$ the sign change lies at $16^\circ$, over the range $13.9$--$14.4\,\mathrm{deg\,d^{-1}}$ the crossing latitude moves from $25^\circ$ to the equator, and at rates above $14.37\,\mathrm{deg\,d^{-1}}$, the equatorial photospheric rate itself, there is no crossing at all and the hole lags at every latitude. The asymmetry in magnitude is secure; the sign reversal at low latitude is not, and should not be used as a diagnostic without an independently determined CH rotation rate.

\section{Wind Signatures and the Acceleration Discrepancy}
\label{sec:wind}

\subsection{Endpoint versus Route} \label{sec:endpoint}

The asymptotic state of a Parker-like wind is a fixed point set by two
integrals of the flow, the mass flux and the total energy per particle. The radial \emph{profile} by which the flow approaches that fixed point is governed instead by where in radius the energy is deposited
\citep{WangSheeley1990,Cranmer2007}. Identical endpoints reached along
different routes are therefore expected when the two sources are structurally distinct.

Figure~\ref{fig:wind} makes this concrete with an illustrative
parameterisation $V(r)=V_\infty[1-e^{-(r-1)/H}]$, $V_\infty=750\,
\mathrm{km\,s^{-1}}$, with $H=3.0\,R_\odot$ (ADCH) and $9.8\,R_\odot$
(EDCH) chosen to reproduce the measured stream, together with mass-flux
conservation and $B\propto r^{-2}$. The $r^{-2}$ scaling is applied only above $R_{ss}$, where PFSS forces the field radial; the faster sub-$R_{ss}$ decline implied by the expansion factors of Section~\ref{sec:expansion} does not enter, and panels of Figure~\ref{fig:wind} should not be read below $\sim2.5\,R_\odot$. We neglect the azimuthal field of the Parker spiral, $B_\phi/B_r=-\Omega r\sin\theta/V$ with $\theta$ the colatitude. It is the one place the rotation axis enters the wind directly, and it differs between the classes, vanishing over the pole but not for equatorial wind; along the equatorial profile of Figure~\ref{fig:wind} it reaches $0.04$ at $10\,R_\odot$ and $0.16$ at $60\,R_\odot$, changing $|\mathbf{B}|$ by at most about $1\%$ inside that radius and leaving the Alfv\'en radius unaffected. Only near 1~au, where it reaches $0.57$, does it matter, raising $|\mathbf{B}|$ by $15\%$ and lowering the equatorial $M_A$ of Figure~\ref{fig:wind}b by about $13\%$ there.
It is anchored on the measured $V(10)=450\,\mathrm{km\,s^{-1}}$,
$n_p(10)=2.5\times10^3\,\mathrm{cm^{-3}}$, and
$V_A(10)=800\,\mathrm{km\,s^{-1}}$, so $M_A(10)=0.56$ is fixed by the inputs rather than predicted; the one quantity carrying independent content is the extrapolated Alfv\'en radius, $R_A=15.6\,R_\odot$, consistent with the measured
$R_A\approx15\,R_\odot$ \citep{Kasper2021,DeForest2016,Finley2025,Badman2026}.

\begin{figure*}
\centering
\includegraphics[width=\textwidth]{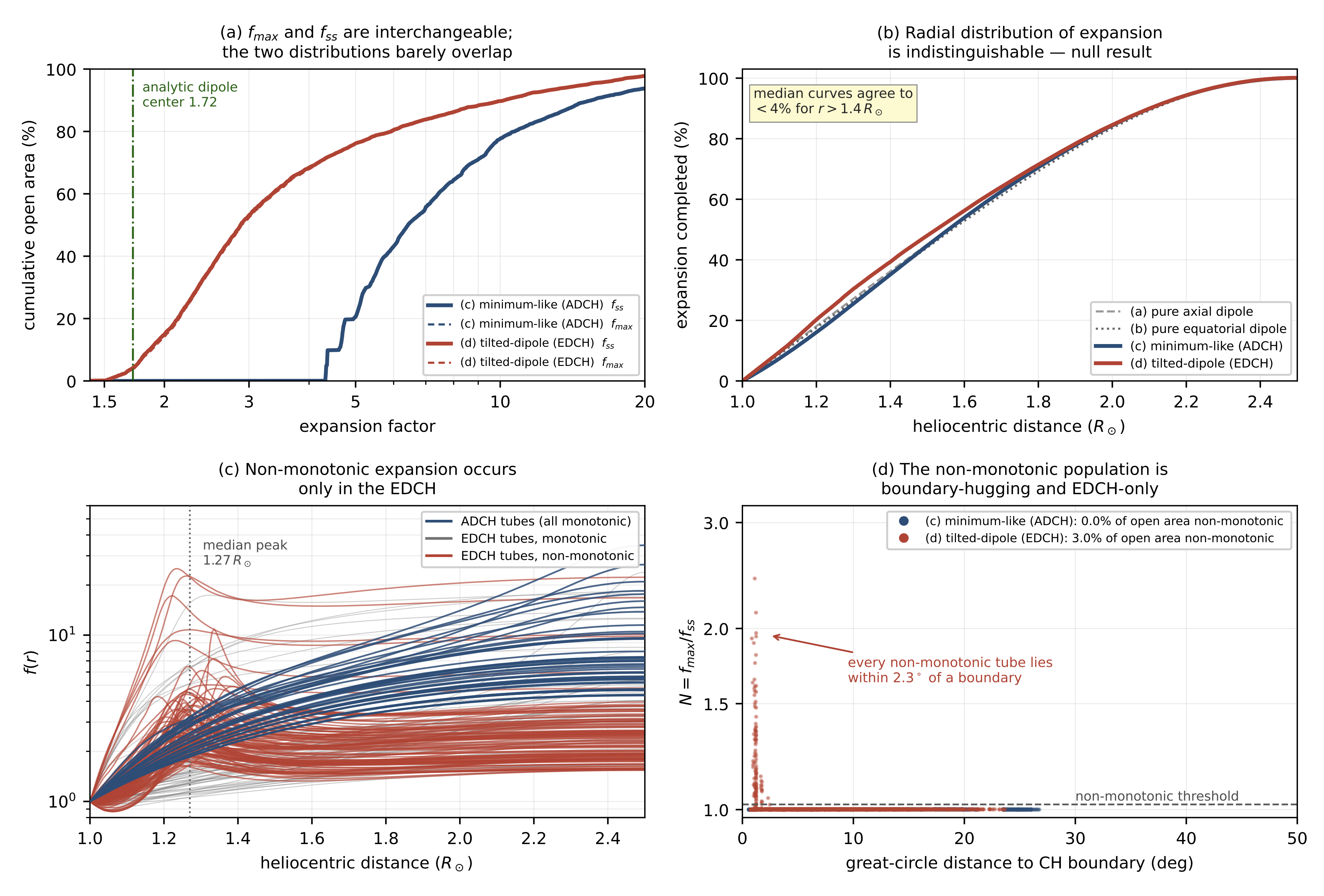}
\caption{Expansion factors along traced open flux tubes from the solutions of
Figure~\ref{fig:maps}, on a grid of 200 radial points, with seeds distributed
uniformly in solid angle so that the congruent cases (a) and (b) are sampled
identically. (a) Cumulative distributions of $f_{ss}$ (solid) and $f_{\max}$
(dashed). The two metrics are interchangeable, differing in the median by
$0.1\%$ for the minimum-like case and $1.0\%$ for the tilted-dipole one. The
two distributions barely overlap: the concentrated polar cap has no tube
expanding by less than $4.3$, while the tilted-dipole hole reaches down to
$1.3$ in its interior.
(b) Completed fraction of the expansion, $[f(r)-1]/[f_{ss}-1]$, as a function
of radius, running from zero at the photosphere to unity at the source surface.
The median curves of all four configurations agree to within $4\%$ above
$1.4\,R_\odot$; the
pure dipoles (a) and (b) lie on top of one another, as congruence requires. (c) Individual $f(r)$ profiles: the non-monotonic tubes
(red), peaking near $1.27\,R_\odot$, occur only in the tilted-dipole case.
(d) Non-monotonicity index $N=f_{\max}/f_{ss}$ against distance to the CH
boundary. $N>1.02$ marks tubes whose expansion peaks below $R_{ss}$; the
threshold lies a factor of thirteen above the numerical floor measured on the
monotonic configurations. Because PFSS forces
$\mathbf{B}\parallel\hat{r}$ at $2.5\,R_\odot$, panel (b) is constrained to
reach $100\%$ there by construction.\label{fig:expansion}}
\end{figure*}

\subsection{Testing the Expansion-Factor Hypothesis}
\label{sec:expansion}

A natural hypothesis is that the equatorial-dipole coronal hole, being a
minority open region embedded in a belt crowded with closed flux and
pseudostreamers, has field lines that remain laterally confined and complete
their expansion at larger radius, so that extended expansion drives extended
acceleration. That expansion factor alone is an incomplete predictor of wind
speed is not itself new: \citet{Riley2001} introduced the distance of an open
footpoint from the nearest coronal-hole boundary as a second parameter,
\citet{Arge2003} combined the two in the WSA model, and \citet{Riley2015}
found the boundary distance to correlate better with 1~au measurements than the
expansion factor does. It is also established that the Wang--Sheeley relation
overestimates the speed of pseudostreamer wind, because pseudostreamer flux
tubes expand non-monotonically with height \citep{Wang2007,RileyLuhmann2012}.
We take these results as established. Our purpose is to isolate them in
configurations where the global field is controlled, and then to test whether
they hold in observed ones. We computed $f(r)=[B(R_\odot)/B(r)]\,(R_\odot/r)^{2}$
along traced open magnetic field lines in each configuration
(Figure~\ref{fig:expansion}). The calculation answers the hypothesis in three
parts. The bulk of the equatorial-dipole hole expands \emph{less} than a polar
hole, at every radius rather than only at the source surface. Normalized by its total, however, the expansion proceeds with height in the
same way in both: by any given height each has completed the same fraction of
its expansion (Figure~\ref{fig:expansion}b), although in absolute terms the
profiles of Figure~\ref{fig:expansion}c differ substantially. And the one genuine
structural difference is topological rather than quantitative: a thin
population of flux tubes along the equatorial-dipole hole's boundary expands
non-monotonically, peaking below the source surface, and has no counterpart in
the axial-dipole configuration. We take the three in turn.

The magnitude of the expansion differs substantially between the two classes,
and in a definite direction. The area-weighted median $f_{ss}$ is $6.42$ at
minimum against $2.88$ in the tilted-dipole configuration, with core values of
$4.36$ and $1.75$, and the two cumulative distributions barely overlap
(Figure~\ref{fig:expansion}a): the minimum-like hole has no tube expanding by
less than $4.3$, being nearly axisymmetric with little interior-to-boundary
contrast, whereas the equatorial-dipole hole reaches down to $1.3$ in its
interior. The difference is not confined to the source surface. Following the
median tube outward, the axial-dipole hole has expanded more at every radius,
by a factor of $1.18$ at $1.1\,R_\odot$, $1.55$ at $1.3\,R_\odot$, $1.86$ at
$1.5\,R_\odot$ and $2.23$ at the source surface. The ordering has a simple
cause: poleward meridional flow concentrates the polar flux and so confines the
boundary of the axial-dipole hole, which must then expand more to fill its
share of the source surface, while the equatorial-dipole hole is under no such
confinement. Its sign is what matters. Larger expansion predicts slower wind,
so if expansion governed the acceleration it would be the \emph{polar} wind
that accelerates more gradually; it is the equatorial-dipole stream that does.
The absolute magnitude of $f_{ss}$ depends on the spatial resolution of the
photospheric map, since higher resolution gives larger footpoint field
strengths $B_0$ while the source-surface field is unaffected, so we do not
compare these values against speed classes calibrated for lower-resolution maps
\citep{WangSheeley2025}. The ordering, computed at a single resolution, is
unaffected by that dependence.

Normalized by its total, by contrast, the expansion proceeds with height in the
same way in every configuration. To separate where a tube expands from how much it expands, we measure the completed fraction $[f(r)-1]/[f_{ss}-1]$, which runs from zero at
the photosphere to unity at the source surface. The ratio $f(r)/f_{ss}$ used in
some treatments begins at $1/f_{ss}$ rather than at zero, so differences in it
near the base restate $f_{ss}$ rather than locating the expansion. The
completed fraction removes the difference in magnitude established above by
design, and so isolates the shape. On this measure the median curves of all
four configurations of Figure~\ref{fig:maps} agree to within $4\%$ at every
radius above $1.4\,R_\odot$; at $1.5\,R_\odot$ the four have completed $45\%$,
$45\%$, $45\%$ and $49\%$ of their expansion (Figure~\ref{fig:expansion}b).
The concentrated polar cap of the minimum-like case, which has by far the
smallest open area, is indistinguishable from the pure dipoles in where its
expansion occurs. Neither the amount nor the radial placement of expansion
therefore gives the equatorial-dipole hole a longer acceleration route: the
first runs the wrong way and the second does not differ. Were extended
expansion the source of extended acceleration, the equatorial-dipole stream
should accelerate at least as quickly as polar wind, whereas it reaches only
$60\%$ of its asymptotic speed at $10\,R_\odot$.

Within the equatorial-dipole configuration the expansion is not uniform.
Splitting its open field into the largest connected component and the thirteen
smaller fragments, which carry half the open area each, gives median $f_{ss}$ of
$2.64$ and $2.98$ respectively. The fragments, the small multipole-associated
holes around the main one, expand more, and they are the population expected to
produce Alfv\'enic slow wind. Both halves nonetheless expand less than the
minimum-like hole, so the ordering above is a property of the configuration as
a whole rather than an artifact of averaging the main hole with its
surroundings.

Because the expansion factor as originally defined \citep{WangSheeley1990}, evaluated between $R_\odot$ and $R_{ss}$, is a poor descriptor of flux tubes
near pseudostreamers, where expansion is non-monotonic with height, rather than increasing steadily as in helmet streamers
\citep{PanasencoVelli2013,Panasenco2019}, we repeated the comparison using the maximum expansion factor along the field line, $f_{\max}$
\citep{WangPanasenco2019}, and the non-monotonicity index
$N \equiv f_{\max}/f_{ss}$.

The comparison survives the change of metric. Area-weighted median $f_{\max}$
is $6.43$ for the minimum-like configuration and $2.90$ for the tilted-dipole
one, against $6.42$ and $2.88$ for $f_{ss}$: the two measures differ by $0.2\%$
and $1.0\%$ respectively and are interchangeable for the bulk of the open area
(Figure~\ref{fig:expansion}a).

The metric does, however, expose a population invisible to $f_{ss}$, and it is
the one structural feature that separates the two classes. Flux tubes with
$N>1.02$, whose expansion peaks below the source surface and then declines,
occupy $3.0\%$ of the open area in the equatorial-dipole hole and none in either
axial-dipole configuration considered here, reaching $N=3.2$ at the extreme
(Figure~\ref{fig:expansion}c,d). The peak lies at $1.19$--$1.50\,R_\odot$
(median $1.27\,R_\odot$), and every such tube lies within $2.3^\circ$ of a CH
boundary. What distinguishes them is topology rather than magnitude: their
median peak expansion, $2.8$, is only $1.2$ times the axial-dipole median at the
same height, although the tail reaches $25$. This is the pseudostreamer-branch
signature identified by \citet{PanasencoVelli2013} and documented with worked
examples for large-scale coronal funnels by \citet{Panasenco2019}; the same
configurations are repeatedly implicated as sources of Alfv\'enic slow wind
\citep{DAmicis2021a,DAmicis2026}. Whether so small an area fraction can carry
the extended acceleration of a bulk stream is an open question we cannot settle
here, but the population identifies where in the hole to look.

The threshold $N>1.02$ is set well above the numerical floor. Configurations (a)
and (b) (Figure~\ref{fig:maps}) expand monotonically by construction, and (c) is
found to do so as well, so any $N>1$ measured there is integration and
interpolation noise: across all $15{,}303$ of their open tubes the largest value
is $N=1.0016$, and the $99.9$th percentile is the same. The cut therefore
exceeds the noise floor by a factor of thirteen, and only twelve tubes in the
tilted-dipole case fall between the floor and the cut, so the classification
does not depend on where within that gap the threshold is placed. Two caveats
bound the claim. First, the absence of the population from the axial-dipole
cases is a property of our chosen minimum-like spectrum, not of solar minima
generally: real polar holes commonly adjoin pseudostreamer lobes, in which
non-monotonic expansion occurs outside reversal phases
\citep{PanasencoVelli2013,Panasenco2019,WangPanasenco2019}. The defensible
statement is that the tilted-dipole configuration makes the boundary
pseudostreamer-dominated, so the population is pervasive there rather than
exclusive to it. Second, the $3.0\%$ figure demonstrates existence rather than
prevalence, and a control calculation with a single compact bipole embedded
inside an open region, the textbook pseudostreamer, yields $N$ up to $6.0$
against $3.2$ here.

To bound how much of the structure of configuration (d) (Figure~\ref{fig:maps}) depends on where its
active-region bipoles happens to sit, we repeated the calculation for $24$
realizations in which the eight bipoles were re-drawn at random latitudes,
longitudes, amplitudes, widths and tilts, holding the equatorial and axial
dipoles fixed. The results are given in the last column of
Table~\ref{tab:metrics}. The open-area fraction is $0.260\pm0.026$, the
maximum boundary distance $32\pm6^\circ$, and the core and median expansion
factors $1.65\pm0.12$ and $2.61\pm0.29$, computed with coarser tracing than the single-realization values of $1.75$ and $2.88$ given above. Two orderings hold in every one of the
$24$ realizations: the tilted-dipole configuration carries more open flux than
the minimum-like one, and its median expansion factor is smaller. Neither is an
artifact of the particular spectrum drawn for Figure~\ref{fig:maps}. The
component count is by far the least stable quantity, $9\pm8$, which is a
further reason to treat it as a property of the model resolution rather than of
the configuration.

The first of those orderings is worth stating plainly, because it is the
observed one and it was not imposed. Only configuration (c) was matched to an
observed open area; the spectrum of configuration (d) was fixed independently,
and it nonetheless carries more open flux, $0.243$ in the realization of
Figure~\ref{fig:maps} and $0.260\pm0.026$ across the ensemble, against $0.137$
at minimum. 
The open flux is expected to be higher in the declining phase than at solar minimum, because the nonaxisymmetric (equatorial dipole) field remains significant while the axial dipole is being rebuilt \citep{WangSheeley2002}; at the surface, this appears as polar holes with prominent equatorward extensions that supply recurrent high-speed streams \citep{WangSheeley2025}.

Interior depth, by contrast, does not separate the classes. The maximum
great-circle distance from an open footpoint to a CH boundary is $24^\circ$ in
the tilted-dipole hole against $28^\circ$ at minimum; once the polar flux is
concentrated as observed, the minimum-like hole is itself compact, and the two
are too close for interior depth to distinguish them. The equatorial-dipole
hole is nonetheless the more fragmented and boundary-dominated of the two, which
is consistent with the distributed interchange reconnection argued for in
Section~\ref{sec:topology}; that is an argument about reconnection and wave
injection rather than about geometric expansion. A systematic survey of expansion profiles across streamer and pseudostreamer sources, distinguishing monotonic from reconverging lanes, is in preparation \citep{PanasencoInPrep}.

This matters because the observed acceleration requires momentum deposition beyond what thermal pressure supplies, implicating fluctuation and switchback momentum flux \citep{Halekas2022,Rivera2024,Squire2020}. Interchange
reconnection at supergranular funnels is an observationally supported source of such fluctuations \citep{Bale2021,Bale2023}. Such reconnection acts low in the corona, but the fluctuations it injects need
not dissipate there: in reflection-driven Alfv\'en-wave turbulence their energy
is deposited over tens of solar radii \citep{Cranmer2007}, the extended heating
that an extended acceleration requires. With the geometric route closed, this is the surviving source-region candidate, but it is a hypothesis about wave driving, and testing it lies outside what a potential-field model can decide. One concrete possibility is that the two classes differ in the purity of their unipolar flux at granular and supergranular scales. The polar cap is close to unipolar, whereas an equatorial-dipole hole rooted in decayed active-region flux retains more mixed polarity, which would alter both the rate of small-scale interchange reconnection at its base and the photospheric
stirring of the network funnels that launches torsional Alfv\'en waves into
the low corona \citep{Finley2022}.

A structural limitation also applies. PFSS forces $\mathbf{B}\parallel\hat{r}$ at $R_{ss}$,
so every solution is constrained to complete its expansion there. The
similarity of the radial profiles is therefore partly an artifact of the
boundary condition, and the confinement hypothesis, which concerns behavior \emph{above} the source surface, is not refuted so much as inaccessible within this framework. Testing it requires an MHD treatment that carries the surrounding streamer and pseudostreamer arcades to well beyond $R_{ss}$.

\begin{figure*}
\centering
\includegraphics[width=0.85\textwidth]{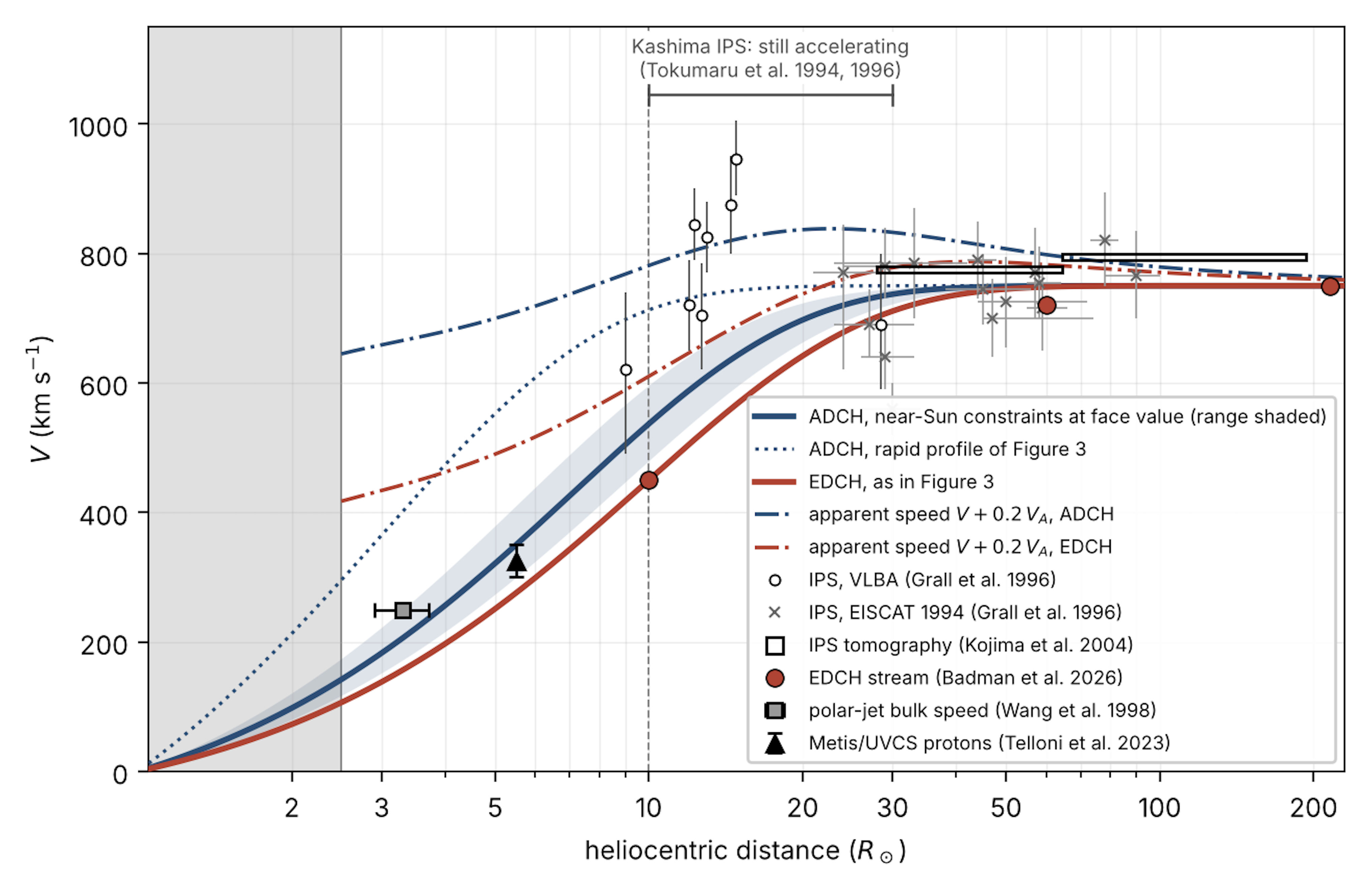}
\caption{Near-Sun constraints on the acceleration of fast wind, drawn with the
parameterization and normalization of Figure~\ref{fig:wind}: a common
$V_\infty=750\,\mathrm{km\,s^{-1}}$, equal mass flux, and the same Alfv\'en
speeds. Grey shading marks $r<R_{ss}$, where the apparent speeds are not drawn.
Solid red: the equatorial-dipole stream of Figure~\ref{fig:wind}
($H=9.8\,R_\odot$), with the measured speeds (filled circles). Dotted blue: the
rapid axial-dipole profile of Figure~\ref{fig:wind} ($H=3\,R_\odot$). Solid
blue with shaded range: axial-dipole profiles passing through the bulk speed of
polar jets at $2.9$--$3.7\,R_\odot$ \citep{Wang1998} and the Metis/UVCS
proton outflow at $5.5\,R_\odot$ \citep{Telloni2023}, both taken at face
value ($H=5.7$--$8.8\,R_\odot$). Dash-dotted: the apparent speed
$V+0.2\,V_A$ of the central axial-dipole profile and of the equatorial-dipole
stream; the factor $0.2$ is chosen for illustration and is not fitted. Also
shown are the VLBA and EISCAT scintillation speeds of \citet{Grall1996}, as in
Figure~\ref{fig:wind}a; the tomographic speeds of \citet{Kojima2004} at
$0.13$--$0.3$ and $0.3$--$0.9$~au (boxes), which refer to the polar wind of
1995--1997 and lie a few per cent above the common $V_\infty$; and the range
over which the Kashima measurements find the wind still accelerating
\citep{Tokumaru1994,Tokumaru1996}.\label{fig:ips}}
\end{figure*}

\begin{figure*}
\centering
\includegraphics[width=0.92\textwidth]{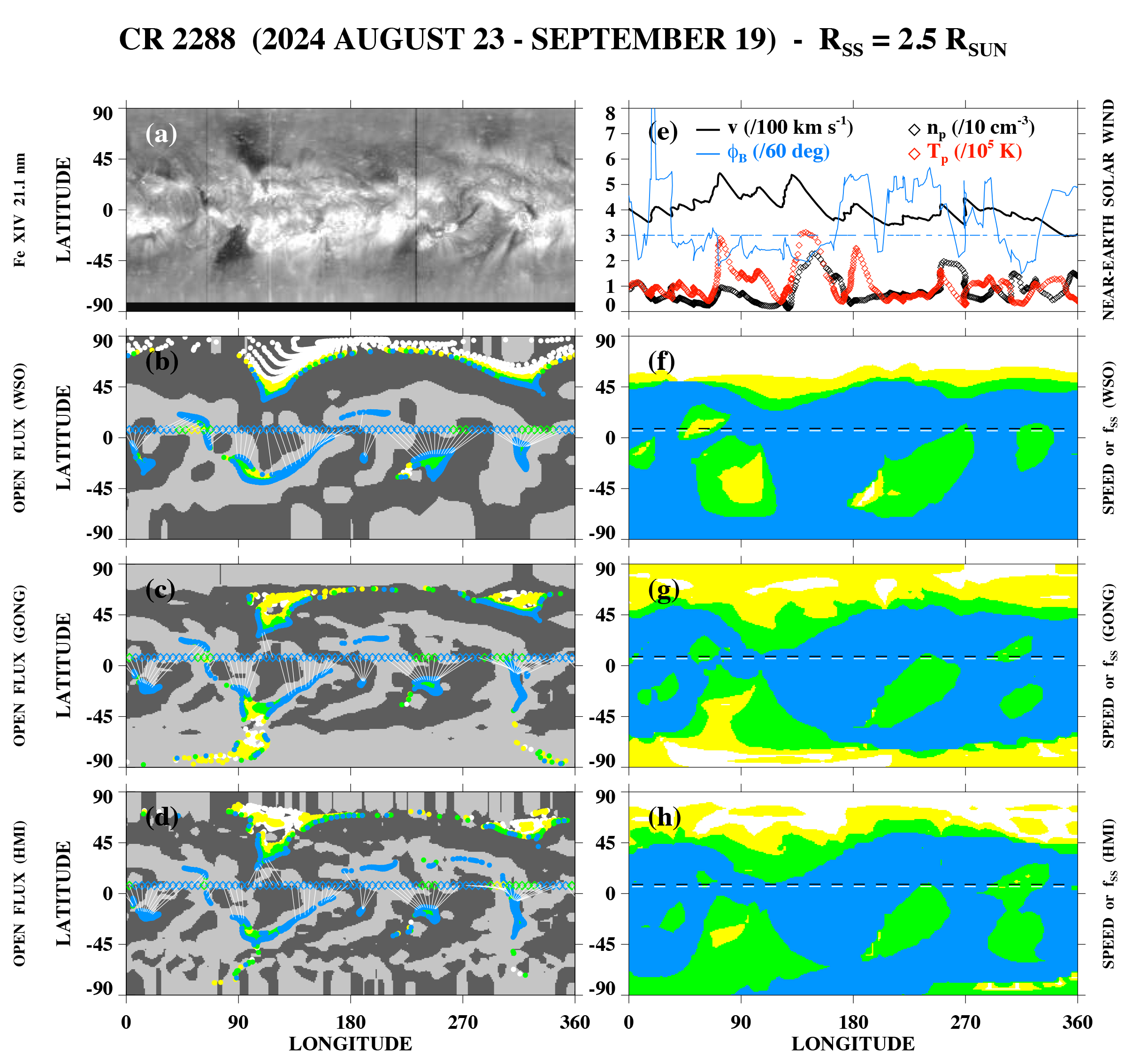}
\caption{Coronal holes and expansion-factor wind-speed predictions for CR~2288
(2024 August 23 to September 19), following the procedure of
\citet[][their Figure~6]{WangSheeley2025}. Panel~(a) shows the
Fe~\textsc{xiv} 21.1~nm emission from SDO/AIA; panels~(b)--(d) the open-field
regions derived from PFSS extrapolations with $R_{ss}=2.5\,R_\odot$ of WSO,
GONG and HMI photospheric maps, colored by the local expansion factor and hence
by the associated wind speed; panel~(e) the near-Earth OMNI measurements \citep{KingPapitashvili2005} of
proton speed, density, temperature and IMF longitude angle; and panels~(f)--(h)
the corresponding distributions of $f_{ss}$ at the source surface, equivalently
the predicted wind speed at 1~au. All maps are smoothed to a common $5^\circ$
resolution before extrapolation, so that the three instruments may be compared
on equal terms. Near the maximum the open flux is confined to small, elongated
unipolar regions and predominantly slow wind is both predicted and observed.
The WSO-derived open regions are systematically larger and less fragmentary
than those from GONG and HMI, a resolution effect discussed in
Section~\ref{sec:obsopen}.
\label{fig:wso2288}}
\end{figure*}

\begin{figure*}
\centering
\includegraphics[width=0.92\textwidth]{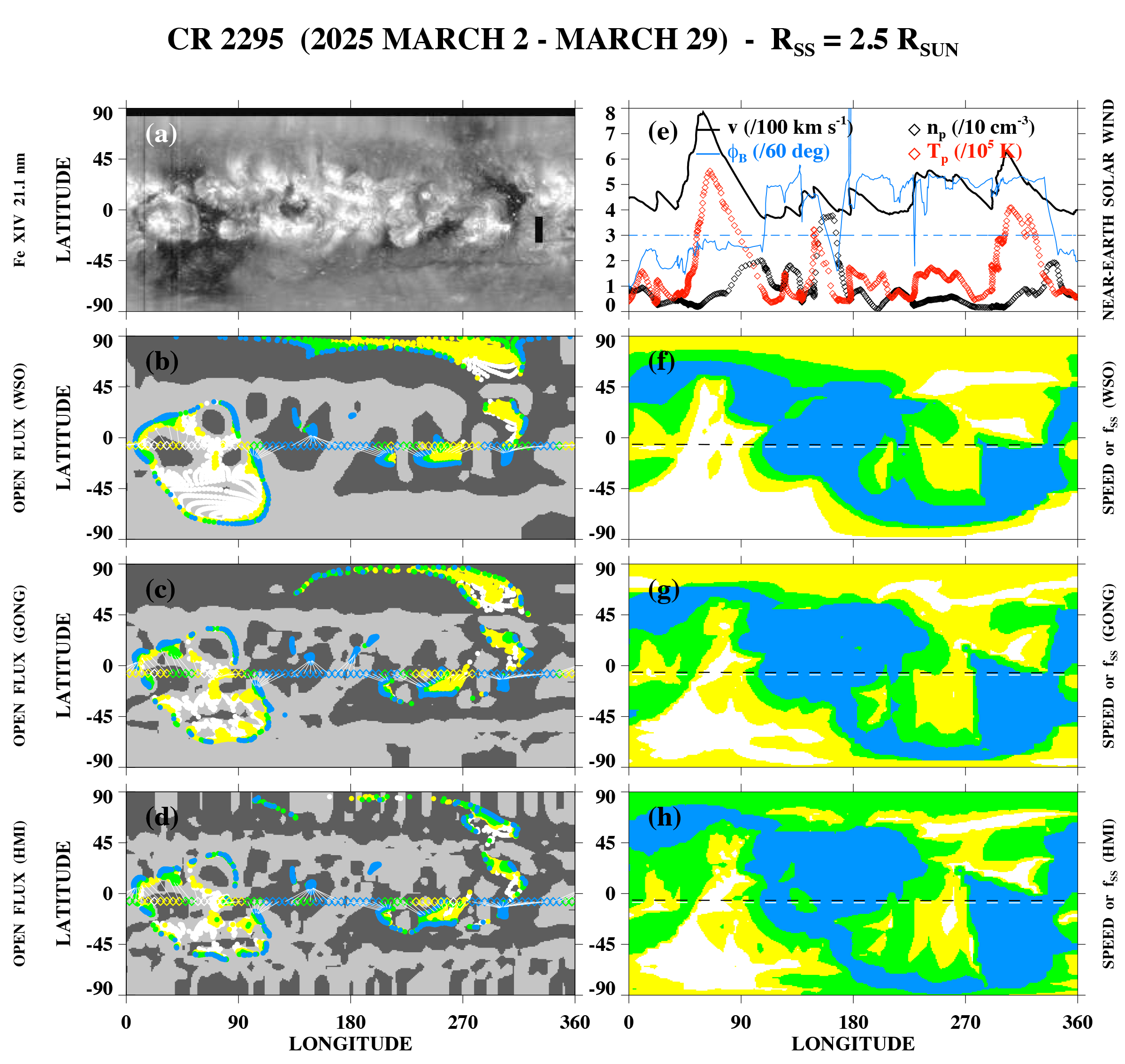}
\caption{As Figure~\ref{fig:wso2288}, for CR~2295 (2025 March 2 to 29) on the
declining phase. A pair of large opposite-polarity equatorial coronal holes has
formed, and fast wind is both predicted and observed.\label{fig:wso2295}}
\end{figure*}

\subsection{A Phase-Speed Bias in the IPS Constraint} \label{sec:ips}

The equatorial-dipole hole is therefore structurally distinct, through its
pseudostreamer-dominated boundary and the continual interchange reconnection it
hosts, and that structure makes it a plausible source of enhanced Alfv\'enic
fluctuation; but it is not distinct in the expansion that would slow its wind
directly. Before attributing the extended acceleration to that structure, it is
worth asking whether the discrepancy it must explain is real, since the polar
side of the comparison rests on scintillation measurements. In deeply sub-Alfv\'enic flow any scintillation tracer propagating at a fluctuation phase speed rather than the bulk speed is biased high by
$(V+V_A)/V$, which for the measured stream is $2.78$ at $10\,R_\odot$. The factor required to reconcile the IPS constraint with $V(10)=450\,
\mathrm{km\,s^{-1}}$ is only $1.67$. Both measurements may therefore be
correct about the quantity they actually sample, with IPS near $10\,R_\odot$ reporting something between $V$ and $V+V_A$ and interpreting it as a completed acceleration \citep{HarmonColes2005}.

The original studies anticipated this possibility. \citet{Grall1996} corrected their speeds for wave bias using an undamped WKB Alfv\'en-wave model, estimated the maximum bias at $\sim15\%$ of the flow speed, and warned that it grows inside $6\,R_\odot$; they could not constrain the fast-wind wave amplitude, their two quoted model fits differing in fast-wind speed by $70\,\mathrm{km\,s^{-1}}$ with wave amplitudes differing by a factor of four, and they closed by calling for direct observations inside $6\,R_\odot$. \citet{Breen1996}, whose own coverage begins at $15\,R_\odot$, stated their conclusion conditionally on the irregularities moving at the bulk speed and noted explicitly that scintillation produced by Alfv\'en waves in an anisotropic medium would not travel at that speed. \citet{Ofman1997}, with Coles, Grall and Klinglesmith among the authors, interpreted the near-Sun scintillation speeds as those of density irregularities moving at speeds between $V$ and $V+V_A$, the band of Figure~\ref{fig:wind}a. Subsequent reanalysis by the same group raised the estimated bias to a substantial fraction of $V_A$ within $\sim20\,R_\odot$ \citep{HarmonColes2005}, and \citet{Kojima2004} note that the large spread of IPS speeds is explained by irregularity motions boosted by wave activity, a spread they find similar in IPS and LASCO measurements at $10$--$20\,R_\odot$. The direct measurement inside the Alfv\'en surface \citep{Badman2026} is the observation \citet{Grall1996} called for.

The IPS evidence for rapid polar acceleration is also less settled than it
first appears. Near-Sun observations at 2 and 8~GHz from Kashima find the wind
still accelerating between about $10$ and $30\,R_\odot$, with high-speed wind
developing over the polar regions as activity declined
\citep{Tokumaru1994,Tokumaru1996}. Tomography of the Nagoya observations places
the high-latitude fast wind at $770$--$780\,\mathrm{km\,s^{-1}}$ between $0.13$
and $0.3$~au, only $19\pm17\,\mathrm{km\,s^{-1}}$ below its speed farther out;
combining this with other measurements, \citet{Kojima2004} conclude that the
fast wind reaches almost its final speed within about $20\,R_\odot$ but
continues to accelerate, slightly but measurably, beyond $30\,R_\odot$, and they
argue explicitly that the acceleration is not complete by $10\,R_\odot$. Those
outer measurements are less affected by the phase-speed bias than the inner
ones, since $V_A$ falls relative to $V$ beyond the Alfv\'en surface, but they are
not immune to it: \citet{Kojima2004} allow a wave bias of about
$130\,\mathrm{km\,s^{-1}}$ at $0.13$~au, falling to $20\,\mathrm{km\,s^{-1}}$ at
$0.9$~au, in which case the true acceleration beyond $0.13$~au exceeds what they
measure. Doppler-dimming measurements of the proton outflow in the polar hole
reach $300$--$350\,\mathrm{km\,s^{-1}}$ at $5.5\,R_\odot$ \citep{Telloni2023};
taken at face value these lie below the rapid profile assumed for
Figure~\ref{fig:wind}a, which gives about $580\,\mathrm{km\,s^{-1}}$ there,
although they sit at the limit of the technique's applicability. White-light
polar jets point the same way. Their leading edges move outward at
$400$--$1100\,\mathrm{km\,s^{-1}}$ while the bulk of their material travels at
about $250\,\mathrm{km\,s^{-1}}$ at $2.9$--$3.7\,R_\odot$, a speed that may
reflect the background polar wind \citep{Wang1998}; in the same polar holes, a
pattern speed exceeds the bulk flow by a factor of two to four. Occasional jets
crossing an IPS line of sight could also contribute high-speed outliers,
although at one to two per day per hole, and followed only to $6\,R_\odot$,
they are too rare to set the average profile.

A difference between the two classes of hole is nonetheless established low in
the corona, by a method the phase-speed bias does not affect. Doppler dimming in
a large equatorial coronal hole near the 1999 maximum gave outflow speeds at
$2$--$3\,R_\odot$ about three times lower than in the polar holes of 1996--1997
at similar heights, although the asymptotic speeds of the two streams differ by
only about $15\%$, so that the bulk of the equatorial-hole acceleration must
occur above $3\,R_\odot$; the equatorial hole was also the denser of the two
\citep{Miralles2001}. What is doubtful is therefore not the contrast itself but
the claim that polar wind completes its acceleration by $10\,R_\odot$. A literal
reading of the IPS profiles, in which fast wind finishes by $10\,R_\odot$ while
slow wind keeps accelerating to several tens of solar radii, has no evident
physical explanation, and the phase-speed bias supplies one for the
\emph{apparent} profile. In low-density fast wind $V_A$ is large, so near
$10\,R_\odot$ the bias is greatest and the sum of a gradually rising $V$ and a
falling $V_A$ can stay close to the terminal speed, mimicking a completed
acceleration; in the denser slow wind $V_A$ is smaller, the bias weaker, and the
gradual rise is seen directly. The contrast IPS reports is thus between fast and
slow wind rather than between polar and equatorial sources, and it may apply
to low-latitude holes too: EISCAT and MERLIN observations near the 1999--2000
maximum found fast flow completing its acceleration inside $10\,R_\odot$
\citep{Breen2002}, flow that \citet{Kojima2004} describe as coming from a large
low-latitude coronal hole.

Figure~\ref{fig:ips} places these constraints on one scale, using the
parameterization of Figure~\ref{fig:wind}. Axial-dipole profiles that pass
through the polar-jet and Doppler-dimming speeds, taken at face value, have
completed only $64$--$80\%$ of their acceleration at $10\,R_\odot$, against
$95\%$ for the rapid profile of Figure~\ref{fig:wind} and $60\%$ for the
equatorial-dipole stream, and they approach their asymptotic speed over
$30$--$60\,R_\odot$, consistent with the continued acceleration beyond
$30\,R_\odot$ found by \citet{Kojima2004}. Adding $0.2\,V_A$ to the central
profile gives an apparent speed that stays between about $700$ and
$840\,\mathrm{km\,s^{-1}}$ from $5$ to $40\,R_\odot$, while the flow itself
accelerates from $320$ to $750\,\mathrm{km\,s^{-1}}$; read literally, it would
place the end of the acceleration near $10\,R_\odot$. The same construction
applied to the denser equatorial-dipole stream gives a weaker bias,
$610\,\mathrm{km\,s^{-1}}$ against a true $450\,\mathrm{km\,s^{-1}}$ at
$10\,R_\odot$, and an apparent profile that still rises visibly beyond that
distance. The factor $0.2$ is chosen for illustration and is not fitted, so the
figure shows how the bias acts but does not measure it. One measurement is not
reproduced. Profiles within the face-value range give an axial- to
equatorial-dipole speed ratio of only $1.1$--$1.6$ at $2.5\,R_\odot$, against
the factor of three found by \citet{Miralles2001}, which only the rapid profile
approaches. We do not resolve this tension here; the single-scale
parameterization is the simplest form consistent with the constraints, and the
true profiles need not follow it.

This is an adequacy argument, not yet a demonstration: the band in
Figure~\ref{fig:wind}a spans $450$--$1250\,\mathrm{km\,s^{-1}}$ and would
accommodate almost any reported value. The discriminating test is the
\emph{radial profile}. Under the bias hypothesis the apparent speed should fall slowly across $10$--$70\,R_\odot$, because $V_A$ declines faster than $V$ rises, whereas the true speed rises monotonically over the same interval. Comparing that predicted shape against published IPS radial profiles, with an explicitly stated tracer weighting, would settle the question. A first step appears in Figure~\ref{fig:wind}a, which overlays the apparent speeds digitized from Figure~3 of \citet{Grall1996}: the VLBA determinations reaching inward of $10\,R_\odot$, the 1994 EISCAT series covering $\sim20$--$100\,R_\odot$ \citep[also shown by][]{Breen1996}, and the density-derived Spartan-201 estimates. Every point lies inside the phase-speed band. The inner VLBA points, at $9$--$15\,R_\odot$, require bias factors of only $1.35$--$1.67$ relative to the extended-acceleration profile, well below the available $(V+V_A)/V\approx2.2$--$3.1$ at those radii, while simultaneously sitting within one standard error of the rapid-acceleration profile with no bias at all. The 1996 dataset is therefore consistent with both readings and cannot discriminate between them; the forward modeling remains the decisive test.

\begin{figure*}
\centering
\includegraphics[width=0.65\textwidth]{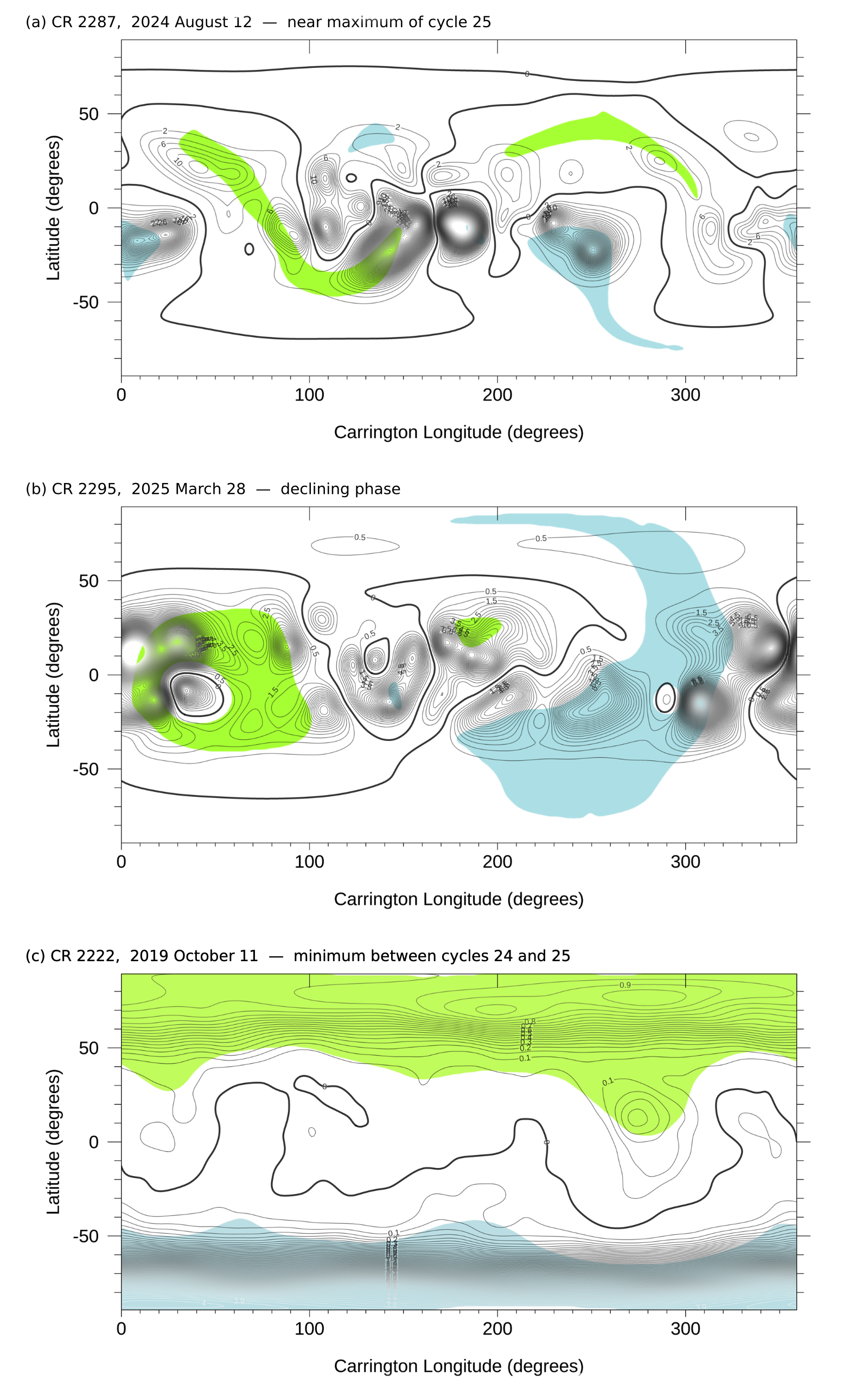}
\caption{Magnetic pressure $B^2$ contour maps at $R=1.2\,R_\odot$ from HMI-driven PFSS
extrapolations with $R_{ss}=2.5\,R_\odot$, for three phases of the cycle.
(a)~CR~2287, 2024 August 12, near the maximum of cycle~25. (b)~CR~2295,
2025 March 28, on the declining phase, the rotation during which the stream of
\citet{Badman2026} was measured. (c)~CR~2222, 2019 October 11, near the minimum
between cycles 24 and 25. Greyscale contours show $B^2$; green and blue shading
marks open field of positive and negative polarity respectively; the heavy
black curve is the neutral line at the same height. The three panels show the sequence the paper is concerned with. At minimum
(c) the neutral line is close to the equator, gently wavy rather than warped,
and the open flux is organized into well-developed polar caps of opposite
polarity: the axial-dipole configuration of Section~\ref{sec:congruence}. Near
maximum (a) no such organization survives; the neutral line is strongly warped
and the open flux is confined to narrow ribbons threaded between active
regions. On the declining phase (b) the open flux appears as two large
regions of opposite polarity, the positive one near Carrington longitude
$50^\circ$ being the source of the fast stream discussed in
Section~\ref{sec:obsdiscrim}. Open-field regions at this height are broader than the photospheric footprints,
since field lines released above the low-lying closed flux reach the source
surface more often; the areas, boundary distances and component counts quoted
in Section~\ref{sec:obsopen} are traced from $r=R_\odot$ and should not be read
off this figure.\label{fig:observed}}
\end{figure*}

Finally, the comparison is unavoidably cross-epoch. The IPS constraints were obtained on polar coronal holes at minimum, when polar magnetic fields, wave energy flux per unit mass flux, and
the proximity of the Alfv\'en surface were near their strongest; at reversal no such hole exists for simultaneous comparison. Part of
the difference in $R_A$ and acceleration length may reflect that contrast
rather than the source-region distinction.

\begin{deluxetable*}{lccccc}
\tablecaption{Open-field metrics for the synthetic configurations of
Figure~\ref{fig:maps} and for a maximum-like sequence in which the coherent
low-order field is progressively suppressed while active-region flux is added.
All rows are computed identically: field lines are traced from the photosphere,
$r=1.0\,R_\odot$, and counted as open if they reach $R_{ss}=2.5\,R_\odot$;
areas are weighted by solid angle; $\theta_b^{\max}$ is the largest
great-circle distance from an open footpoint to a coronal-hole boundary
\citep{Riley2001}; and the expansion is carried to $\ell_{\max}=30$. Each entry
is a single realization of the corresponding spectrum. The final row gives the
mean and standard deviation over $24$ realizations of configuration (d) in
which the eight active-region bipoles are re-drawn at random.\label{tab:metrics}}
\tablehead{\colhead{Configuration} & \colhead{open area} &
\colhead{$\theta_b^{\max}$ (deg)} & \colhead{components} &
\colhead{largest} & \colhead{$l=1$ power}}
\startdata
\multicolumn{6}{l}{\textbf{Configurations of Figure~\ref{fig:maps}}} \\
\hline
(a) pure axial dipole             & 0.354 & 46 &   2 & 0.177 & 100\% \\
(b) pure equatorial dipole        & 0.353 & 46 &   2 & 0.177 & 100\% \\
(c) minimum-like (ADCH)           & 0.137 & 28 &   2 & 0.068 & 49.4\% \\
(d) tilted-dipole (EDCH)          & 0.243 & 24 &  19 & 0.122 & 11.9\% \\
(e) tilted dipole $+$ axial quadrupole & 0.211 & 22 &  17 & 0.101 &  0.4\% \\
(f) maximum-like ($=$f5)          & 0.012 &  3 &  62 & 0.003 &  0.1\% \\
\hline
\multicolumn{6}{l}{\textbf{Maximum-like sequence}} \\
\hline
f1 (12 ARs, dipole 3.0)           & 0.284 & 36 &  13 & 0.156 &  4.3\% \\
f2 (18 ARs, dipole 1.5)           & 0.157 & 21 &  22 & 0.079 &  0.8\% \\
f3 (24 ARs, dipole 0.8)           & 0.103 & 19 &  48 & 0.060 &  0.2\% \\
f4 (30 ARs, dipole 0.4)           & 0.040 &  4 &  66 & 0.018 &  0.1\% \\
f5 (38 ARs, dipole 0.2)           & 0.012 &  3 &  62 & 0.003 &  0.1\% \\
\hline
\multicolumn{6}{l}{\textbf{Ensemble over bipole placements, configuration (d)}} \\
\hline
24 realizations, mean $\pm$ s.d. & $0.260\pm0.026$ & $32\pm6$ & $9\pm8$ & $0.145\pm0.017$ & \nodata \\
\enddata
\tablecomments{The sequence f1--f5 suppresses the coherent low-order field
while raising the active-region flux, and moves the open field from a coherent
regime, with a few large components and interiors reaching several tens of
degrees, to a fragmented one in which every open footpoint lies within a few
degrees of a boundary. Open area, $\theta_b^{\max}$ and the component count all
depend on the harmonic truncation and on the radius at which field lines are
released, so values here are comparable with one another but not with those
derived from lower-resolution synoptic maps; see Section~\ref{sec:obsopen}.
Power fractions in the final column use a denominator summed over $l\le20$.}
\end{deluxetable*}

\section{Observed Configurations} \label{sec:observed}

The synthetic spectra of Section~\ref{sec:models} are idealized. To establish
that they correspond to configurations the Sun actually adopts, and to test
the structural distinctions of Section~\ref{sec:structure} against data, we
applied the identical analysis to two PFSS extrapolations computed from
HMI synoptic magnetograms with the SolarSoft package \citep{Schrijver2003},
using the same source surface $R_{ss}=2.5\,R_\odot$. The SolarSoft solution is
evaluated on its native grid of $384\times192\times39$ points, uniform in
longitude at $0\fdg9375$, on Gauss--Legendre nodes in $\sin(\mathrm{latitude})$,
and logarithmically spaced in radius from $1.0$ to $2.5\,R_\odot$, with
harmonics retained to $\ell_{\max}=192$. Field lines are traced from every
photospheric pixel using the package's own integrator, and classified as open
by the same test applied in Section~\ref{sec:models}: a line is open if it
reaches $R_{ss}$.
Two properties of the input maps should be stated, because they differ from
those used in the comparison of Figure~\ref{fig:wso2288} and in the wind-speed
predictions of \citet{WangSheeley2025}. First, ours are instantaneous
full-surface maps at 12-hour cadence, combining the observed near side with a
reconstructed far side, rather than $27$-day Carrington synoptic maps assembled
from successive central-meridian strips. We select single days rather than
whole rotations because the configuration changed rapidly over the interval of
interest: a synoptic map spanning 2024 August 23 to September 19 averages over
part of that change, whereas a single-day map characterizes the field at the
sunspot maximum itself. This is why our maximum-phase date falls in CR~2287
while the synoptic comparison in Figure~\ref{fig:wso2288} uses CR~2288.
Second, ours are used at their native resolution, $\ell_{\max}=192$, whereas
\citet{WangSheeley2025} smooth all maps to a common $5^\circ$ pixel before
extrapolating, so that expansion factors derived from different observatories
may be compared on equal terms. Open-field areas and expansion factors depend
on the resolution of the input field \citep[see also][]{Poduval2020}, and
values from the two approaches are not directly comparable.

The far-side contribution is modelled rather than observed, and it enters the
low-order harmonics discussed in Section~\ref{sec:obsharm}. For the March
rotation this
carries little weight for the question at issue: the coronal hole that produced
the fast stream lay on the near side, and Earth was magnetically connected to
it on the date analyzed.

Figure~\ref{fig:observed} displays the magnetic pressure $B^2$ at
$1.2\,R_\odot$ rather than the photospheric field itself. At that height the
low-lying closed flux of the active regions has largely fallen away, and the
large-scale structure that organizes the corona is easier to see; the
photospheric field of the same rotations appears instead in the upper panels of
Figure~\ref{fig:wso2288}. The open-field shading of
Figure~\ref{fig:observed} is likewise evaluated at $1.2\,R_\odot$, and the
regions it marks are correspondingly broader than the photospheric footprints:
field lines released above the closed flux reach the source surface far more
often than those released at $r=R_\odot$. All open-field areas, boundary
distances and component counts quoted in Section~\ref{sec:obsopen} are traced
from the photosphere, and should not be read off Figure~\ref{fig:observed}.
The two dates are
CR~2287 (2024 August 12), close to the sunspot maximum of cycle~25, and
CR~2295 (2025 March 28), on the declining phase and the rotation during which
the stream of \citet{Badman2026} was measured.
Both are shown in Figure~\ref{fig:observed}.

The observed extrapolations differ from the synthetic solutions in two
respects. The first is the harmonic truncation. We retain
$\ell_{\max}=22$--$30$ in the idealized configurations, where the low-order
field is imposed by construction and the higher harmonics carry little power,
whereas the observed extrapolations retain $\ell_{\max}=192$: $94$--$96\%$ of
the power in both observed rotations lies at $\ell\ge4$
(Section~\ref{sec:obsharm}), against $74\%$ in our tilted-dipole spectrum, so
a truncation adequate for the synthetic fields would discard most of an
observed one. The synthetic spectra should accordingly be read as controlled
simplifications, not as representative realizations of the photospheric field
at these phases.

\begin{deluxetable}{lrr}
\tablecaption{Low-order harmonic content of the two observed rotations, from
exact Gauss--Legendre quadrature on the native grid of the extrapolations.
Coefficients are Schmidt semi-normalized and in gauss; $|g_{lm}|$ denotes the
quadrature sum of the two $m\ne0$ coefficients. Power fractions are relative to
the total in $l\le3$.\label{tab:harmonic}}
\tablehead{\colhead{Component} & \colhead{CR 2287} & \colhead{CR 2295}}
\startdata
$g_{10}$\ \ axial dipole      & $+0.29$ & $+0.35$ \\
$|g_{11}|$\ equatorial dipole &  1.05   &  2.15   \\
$g_{20}$\ \ axial quadrupole  & $+0.27$ & $-0.89$ \\
$|g_{21}|$                    &  2.47   &  1.11   \\
$|g_{22}|$                    &  1.90   &  3.35   \\
\hline
power in $l=1$                &  4\%    & 27\%    \\
power in $l=2$                & 19\%    & 45\%    \\
power in $l=3$                & 77\%    & 28\%    \\
\enddata
\tablecomments{In CR~2287 the equatorial dipole is not the largest low-order
component: both $|g_{21}|$ and $|g_{22}|$ exceed it, and $l=3$ carries most of
the power below $l=4$. In CR~2295 the equatorial dipole has doubled and the
$l=1$ power has risen by a factor of seven, although $|g_{22}|$ still exceeds
$|g_{11}|$.}
\end{deluxetable}

The second concerns what is compared. Although the two sets of field solutions
are obtained by different codes and at different truncations, every derived
quantity reported in Table~\ref{tab:metrics} is computed from the
resulting open-field mask, using the same definitions and the same solid-angle
weighting in both cases: the open-area fraction as the solid-angle-weighted
fraction of the photosphere from which field lines reach $R_{ss}$;
$\theta_b^{\max}$ as the largest great-circle distance from an open footpoint
to the nearest coronal-hole boundary \citep{Riley2001}; and the component
count from a connected-component decomposition of the open mask with longitude
treated as periodic. The entries of Table~\ref{tab:metrics} are therefore
comparable because the post-processing is common, even though the field
solvers are not.

\subsection{Harmonic Content} \label{sec:obsharm}

Table~\ref{tab:harmonic} gives the low-order content of each rotation, obtained
by decomposing the photospheric $B_r$ on its native Gauss--Legendre grid.
Neither rotation is dominated by the equatorial dipole. In CR~2287 the
equatorial dipole $|g_{11}|=1.05$~G is exceeded by both $|g_{21}|=2.47$~G and
$|g_{22}|=1.90$~G, and $l=3$ carries $77\%$ of the power below $l=4$: the open
flux at that time is associated with the quadrupole and higher multipoles
rather than with the equatorial dipole, and the holes of CR~2287 are
accordingly \emph{not} equatorial-dipole coronal holes. In CR~2295 the
equatorial dipole has doubled to $2.15$~G and the $l=1$ power has risen from
$4\%$ to $27\%$, so that configuration is much closer to the canonical tilted
dipole, although $|g_{22}|=3.35$~G still exceeds $|g_{11}|$. The axial dipole
is small in both, $+0.29$ and $+0.35$~G, as expected of a field that has
recently reversed and not yet rebuilt its polar caps.

This ordering is the expected one. When the axial dipole passes through zero
the equatorial dipole need not take over: the $(2,2)$ quadrupole and other
multipoles can carry much of the field, and commonly become important first.
The axial dipole then re-emerges with reversed sign, and a tilted dipole may
follow \citep{Wang2014}, although whether a clear one forms depends on how the
emerging flux is distributed. In cycle~25 one did, once the nests of 2025
January had emerged \citep{Finley2026}.
The equatorial-dipole coronal hole of Section~\ref{sec:congruence} is
therefore a configuration of the interval in which the tilted dipole is
established, not of the reversal itself.

Two features deserve comment. First, $94$--$96\%$ of the power in both
rotations lies at $l\ge4$, against $74\%$ in our tilted-dipole synthetic
spectrum; real fields at these phases are considerably less dipolar than any
of our idealized cases. Second, the mean
photospheric $B_r$ poleward of $60^\circ$ is negative in \emph{both}
hemispheres in both rotations, $-0.08$ and $-0.57$~G for CR~2287 and $-1.68$
and $-0.93$~G for CR~2295. Same-signed polar fields correspond formally to a polar signal carried by the
axial quadrupole rather than the axial dipole, the component that hemispheric
asymmetries in sunspot activity generate and that displaces the heliospheric
current sheet north--south \citep{Wang2014}, and they motivate the quadrupolar
configuration (e) of Figure~\ref{fig:maps} and Table~\ref{tab:metrics}. The axial quadrupole
$g_{20}$ is $+0.43$ in CR~2287 and $-0.89$~G in CR~2295, the latter more than
twice the axial dipole.

That description should be treated with care, for two reasons. First, our polar
values come from extrapolations whose polar caps are filled by a model rather
than observed, and the coefficients most sensitive to that filling are exactly
the axisymmetric ones. Second, the first direct out-of-ecliptic measurements,
made by the Polarimetric and Helioseismic Imager on board Solar Orbiter (SO/PHI-HRT) in March and April 2025 and therefore contemporaneous with
CR~2295, give net flux densities above $70^\circ$ of $-0.66\pm0.32$ and
$+0.31\pm0.25\,\mathrm{Mx\,cm^{-2}}$ for the north and south poles
respectively \citep{Calchetti2025}. Those are of opposite sign, whereas our
extrapolation gives the same sign at both poles. The southern value is only
$1.2\sigma$ from zero, and above about $85^\circ$ the measured net flux
approaches zero at both poles, so the two hemispheres are not clearly
distinguished in the polar-most bins. We therefore read the same-signed result
as a statement about the low-order decomposition of a fragmented,
largely cancelled polar field rather than as evidence for a coherent
quadrupolar cap.

\begin{figure*}
\centering
\includegraphics[width=0.8\textwidth]{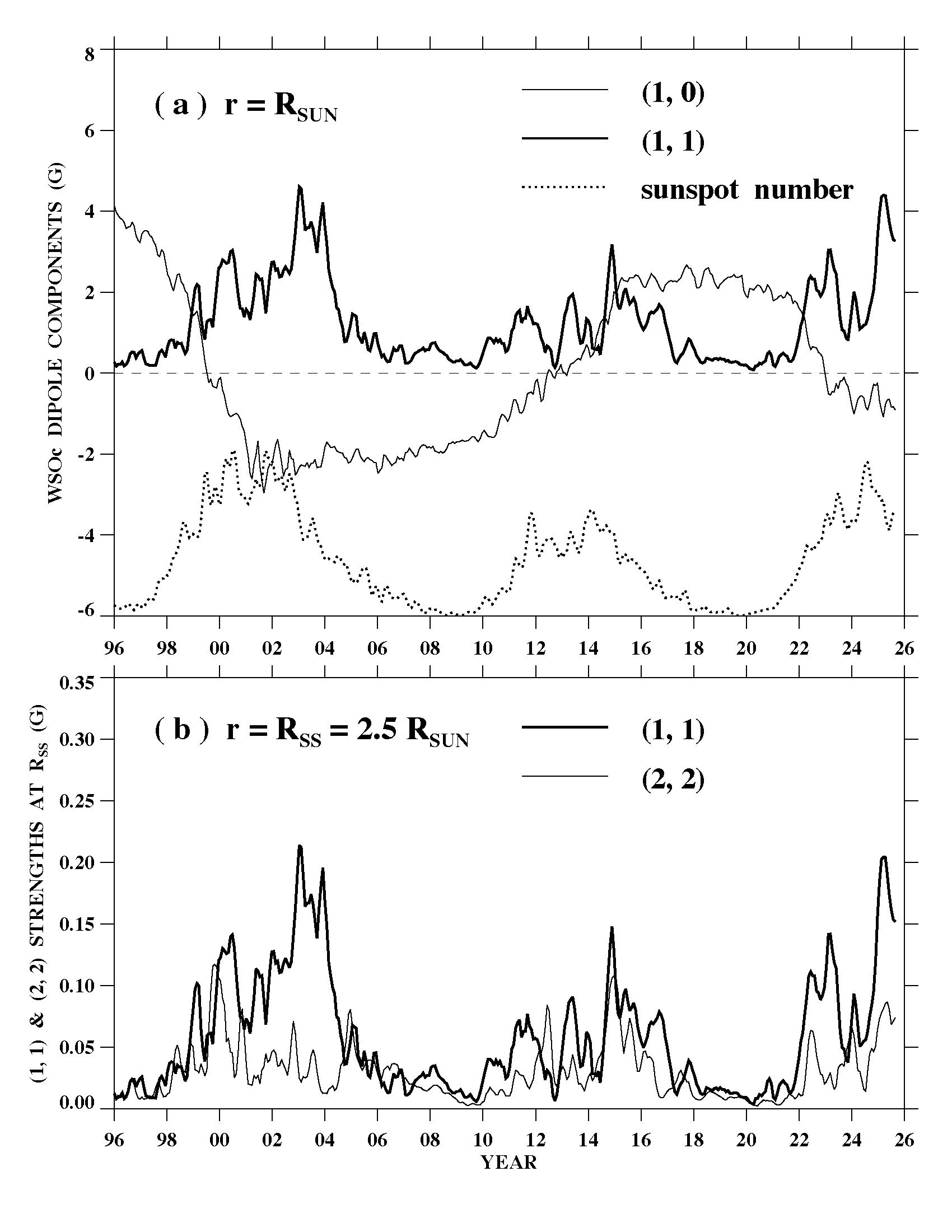}
\caption{Low-order harmonic content of the coronal field over three cycles,
from WSO measurements. (a) Photospheric $(1,0)$ and $(1,1)$ dipole components
(G), with the sunspot number for phase reference. (b) $(1,1)$ and $(2,2)$
strengths evaluated at the source surface $R_{ss}=2.5\,R_\odot$, where the
high harmonics have been filtered away. Both panels are 3-CR running averages
of PFSS extrapolations of WSO photospheric maps, running from CR~1904 (1995
December 20) to CR~2302 (2025 September 9 -- October 6), the last complete WSO
map available. The saturation correction of \citet{WangSheeley1995} has been
applied. Quadrupole-led
intervals occur near the activity maxima, including mid-2012
(Section~\ref{sec:obsharm}); strong $(1,1)$ intervals occur on the declining
phases and are in general accompanied by large recurrent high-speed streams.
The 2025 rise of the $(1,1)$ component separates CR~2287 from CR~2295.
\label{fig:multipoles}}
\end{figure*}

The dominance of a non-dipolar component at these phases has precedent.
During the cycle 24 reversal in mid-2012 the nonaxisymmetric $(2,2)$
quadrupole became the largest low-order component, and the heliospheric
current sheet split into two cylindrical sheets, each centered on a
low-latitude coronal hole of the same polarity \citep{Wang2014}. Those holes
are, in the sense of Section~\ref{sec:congruence}, footprints of a quadrupole
rather than of the equatorial dipole, and the taxonomy introduced here
extends to them directly: the harmonic component responsible is again the
observable that constrains the configuration's geometry and lifetime. The
equatorial-dipole hole is the $l=1$ case of the same construction.

The rise in $|g_{11}|$ between our two rotations has an identified cause.
\citet{Finley2026}, tracking active nests through the maximum of cycle~25,
find that two nests emerged from a collision site in 2025 January and a third
appeared in the northern hemisphere, and that this emergence reinforced a
tilted-dipole topology which persisted through 2025; \citet{Badman2026} reach
the same attribution for the source of their stream. The surge is therefore
traceable to specific flux emergence at specific longitudes rather than to a
gradual global evolution, consistent with the amplitude behavior of
Section~\ref{sec:residence}, and it moves the corona along the f1--f5 sequence
of Table~\ref{tab:metrics}, whose control parameter is exactly the amplitude of
the coherent component relative to the active-region flux. Across that sequence
the open area falls from $0.28$ to $0.01$ and the largest boundary distance from
$36^\circ$ to $3^\circ$, as the open field fragments into many small regions;
adding an axial quadrupole to the tilted dipole, configuration (e), changes the
same metrics only modestly, from $0.24$ to $0.21$ in open area and from
$24^\circ$ to $22^\circ$ in boundary distance. The same alternation
recurs over three cycles (Figure~\ref{fig:multipoles}). At the source surface,
quadrupole-led intervals near the activity maxima, including the mid-2012
episode above, alternate with declining-phase intervals in which the equatorial
dipole dominates, and a strong $(1,1)$ component is in general accompanied by
large recurrent high-speed streams \citep{Ebert2009,Gyeltshen2026}. The 2025
surge is one instance of that pattern.

\subsection{Open-Field Structure} \label{sec:obsopen}

Tracing a field line from every photospheric pixel and classifying it as open
if it reaches $R_{ss}$ gives open-area fractions of $0.016$ for CR~2287 and
$0.064$ for CR~2295. We attach little weight to these numbers, and none to a
comparison between them, for three reasons.

First, the open-area fraction depends strongly on the harmonic truncation of
the input field. Repeating the extrapolation and the tracing with the same
solutions truncated at $\ell_{\max}=9$ and $20$ gives $0.091$ and $0.065$ for
CR~2287 and $0.106$ and $0.073$ for CR~2295, a factor of six between
$\ell_{\max}=9$ and the full $\ell_{\max}=192$, because fine-scale flux
resolved at high degree closes locally instead of opening to the source
surface. Open-area series built from WSO synoptic maps, which retain harmonics
only to $\ell\approx9$ and give of order $5\%$ at solar maximum
\citep{Wang2009}, are therefore not directly comparable with values from
HMI-driven extrapolations. The number of disconnected open regions behaves the
same way, ranging over more than an order of magnitude across the same
truncations, and is not a well-defined property of a configuration.

Second, the footprint depends on the height at which field lines are released.
Seeding at $1.2\,R_\odot$ rather than at the photosphere raises the open area
to $0.262$ and $0.290$ for the two rotations, because lines released above the
low-lying closed flux reach the source surface far more often.

Third, and specific to these rotations, the source of the fast stream is a
positive-polarity region near Carrington longitude $50^\circ$ adjoining an
active-region nest. At $1.2\,R_\odot$ that region is broad and coherent and is
plainly the stream source; at the photosphere only a tenth of it is open in our
extrapolation, the open flux emerging from scattered footpoints between the
active-region flux. Whether that reflects a genuinely narrow funnel or the
failure of a high-degree potential-field solution to keep flux open beside
strong active-region fields, we cannot determine here. Non-monotonic expansion
in funnels of this kind \citep{Panasenco2019} would in any case make the
expansion factor evaluated between $R_\odot$ and $R_{ss}$ a poor descriptor of
the flux tubes involved, as Section~\ref{sec:expansion} discusses.

Taken together, these sensitivities mean that our extrapolations do not
establish a difference in open-field organization between the two rotations,
and we claim none. The comparison these data do support is the harmonic one of
Section~\ref{sec:obsharm}.

\subsection{The Parker Solar Probe Stream from CR 2295} \label{sec:psp}

The equatorial-dipole hole of CR~2295 was crossed by Parker Solar Probe during
its twenty-third perihelion, at a closest approach of $9.86\,R_\odot$, and the
same stream was measured by Solar Orbiter near $60\,R_\odot$ and by Wind at
L1 \citep{Badman2026}. Ballistic mapping of all three spacecraft to
$2.5\,R_\odot$, followed by PFSS tracing to the photosphere, places their
footpoints within the same coronal hole and close enough to one another to
appear as a single track, with each crossing passing through a local maximum in
wind speed at the most interior footpoint it samples. The hole occupies $7\%$
of the solar surface, comparable to the $6$--$10\%$ of a solar-minimum polar
hole \citep{deToma2011}.

At $10\,R_\odot$ the stream is deeply sub-Alfv\'enic, $M_A\approx0.6$, with a
bulk speed of $468^{+26}_{-41}\,\mathrm{km\,s^{-1}}$, an Alfv\'en speed near
$800\,\mathrm{km\,s^{-1}}$, a proton density of
$2900^{+200}_{-300}\,\mathrm{cm^{-3}}$, and a proton temperature of
$2.58^{+0.38}_{-0.40}$~MK against an electron temperature near $0.8$~MK
\citep{Badman2026}. The inferred Alfv\'en critical point lies at
$15.2\,R_\odot$. By 1~au the stream reaches
$763^{+11}_{-11}\,\mathrm{km\,s^{-1}}$ at a density of
$2.1^{+0.2}_{-0.2}\,\mathrm{cm^{-3}}$, values indistinguishable from the polar
flows characterized by Ulysses \citep{Phillips1995,McComas2000}. The speed at
$10\,R_\odot$ is therefore only about $60\%$ of the asymptotic value, one third
of the acceleration occurs beyond $10\,R_\odot$, and $95\%$ of it is complete
by $60\,R_\odot$. Protons and electrons cool along a single power law through
all three measurement points, with polytropic indices $\gamma_p=1.35$ and
$\gamma_e=1.31$.

In the taxonomy of this paper the source is an equatorial-dipole hole.
\citet{Badman2026} note that the hole was aligned with the solar magnetic
dipole axis, which dominated the coronal structure at the time, and trace that
alignment to the inclined dipolar field that persisted for several months
following the nesting of active regions in 2024 \citep{Rivera2025b, Finley2026}.
An inclined dipole whose axis lies near the equatorial plane is precisely the
$g_{11}$-dominated configuration of Section~\ref{sec:congruence}, and the surge
in $|g_{11}|$ documented in Section~\ref{sec:obsharm}, from $1.05$ to $2.15$~G
between CR~2287 and CR~2295, is the same event seen in the harmonic
decomposition. The stream is thus the wind of an EDCH sampled at three
heliocentric distances. What distinguishes this rotation from CR~2287, which
produced no comparable stream, is the low-order field rather than any robustly
measured difference in open-field organization: the $l=1$ power had risen from
$4\%$ to $27\%$ and the corona had passed from quadrupole-led to dipole-led
(Section~\ref{sec:obsharm}).

Two points of contact with the preceding sections deserve comment. First, the
measured Alfv\'en critical point at $15.2\,R_\odot$ agrees closely with the
$15.6\,R_\odot$ returned by the parameterisation of Section~\ref{sec:endpoint},
which was anchored only on $V$, $n_p$ and $V_A$ at $10\,R_\odot$ and carries
$R_A$ as its one genuinely extrapolated quantity. Second,
\citet{Badman2026} reach the same reading of the scintillation discrepancy
developed in Section~\ref{sec:ips}, and from the opposite direction: they note
that $V_R+V_A$ for this stream is $1250$--$1300\,\mathrm{km\,s^{-1}}$ at
$10\,R_\odot$, above the largest apparent speeds reported by
\citet{Grall1996}, and argue that a systematic overestimate from the convolution
of bulk and wave speeds is more likely than a genuine difference between polar
and equatorial acceleration profiles. Our Figure~\ref{fig:wind}a shows the same
thing geometrically: every published scintillation point lies inside the band
bounded by $V$ and $V+V_A$.

Finally, \citet{Badman2026} find the fluctuations at $10\,R_\odot$ to be
spherically polarized but of small deflection angle, $20$--$30^\circ$, becoming
full-reversal switchbacks by $60\,R_\odot$, and estimate the Alfv\'en wave
energy density at $45\%$ of the proton kinetic energy density at Parker,
sufficient to supply roughly $100\,\mathrm{km\,s^{-1}}$ of the acceleration
budget if fully converted. This is the wave-driving channel identified in
Section~\ref{sec:expansion} as the surviving source-region candidate once the
geometric route is closed, now with a measured amplitude rather than an
inferred one.

\subsection{Is the March 2025 Hole Polar-like?} \label{sec:obsdiscrim}

CR~2295 produced a stream reaching $750$--$800\,\mathrm{km\,s^{-1}}$
\citep{Badman2026}, while OMNI measurements \citep{KingPapitashvili2005} through CR~2287 show
speeds that
barely exceed $500\,\mathrm{km\,s^{-1}}$. That contrast is typical of the
phases involved. Ulysses established that the wind is sharply bimodal at
minimum, with fast coronal-hole flow at high latitude and slow streamer-belt
flow near the equator, and that this organization breaks down at maximum, when
slow and fast wind are interleaved without clear latitudinal order
\citep{McComas2000,McComas2003,Ebert2009}. Modeling the ambient wind over four
and a half cycles, \citet{Gyeltshen2026} find both the mean equatorial speed
and its variability to be systematically lower at maximum than at minimum, with
the declining phase carrying the highest short-term variability of any phase.

By its asymptotic properties the CR~2295 stream is polar-like, and
\citet{Badman2026} establish this in detail: $763\,\mathrm{km\,s^{-1}}$,
$2.1\,\mathrm{cm^{-3}}$ and $0.36$~MK at 1~au, all matching the Ulysses polar
values, from a hole covering $7\%$ of the surface against $6$--$10\%$ for a
solar-minimum polar hole \citep{deToma2011}.

By its magnetic configuration it is not. The congruence of
Section~\ref{sec:congruence} holds for a \emph{pure} equatorial dipole, and
CR~2295 is not one: $|g_{22}|$ still exceeds $|g_{11}|$, and $l=2$ carries more
power than $l=1$.
It is a tilted dipole emerging from a quadrupole-led corona, with its principal
open region adjoining an active-region nest, rather than a dipole pole
displaced to the equator.

We regard this asymmetry as a result rather than a contradiction to be
resolved. The asymptotic wind state does not diagnose the harmonic
configuration of its source: a coronal hole need not be a dipole pole in order
to produce wind indistinguishable from polar flow. The congruence argument
identifies when two holes are magnetically the same object, and by that
criterion the March 2025 hole is not a polar hole, notwithstanding its wind.
Descriptions such as ``polar-hole-like'' are therefore best read, as
\citet{Badman2026} intend them, as statements about the outflow rather than
about the field that produced it.

What separates the two rotations in the wind is the organization of the open
flux, although our own extrapolations cannot establish that contrast robustly.
At the common $5^\circ$ resolution of Figures~\ref{fig:wso2288}
and~\ref{fig:wso2295}, the open flux near maximum lies in small, elongated
unipolar regions and predominantly slow wind is both predicted and observed,
whereas on the declining phase a pair of large equatorial holes has formed and
fast wind is both predicted and observed; at the native resolution of our
HMI-driven extrapolations the contrast is not recovered robustly
(Section~\ref{sec:obsopen}). The boundary-distance and expansion-factor
descriptions of that organization are two views of one structure: rapidly
expanding tubes are rooted near coronal-hole boundaries and slowly expanding
ones far from them, so boundary distance \citep{Riley2001,Riley2015} inherits
its predictive power from the expansion it proxies \citep{WangSheeley1990}, and
a configuration whose open flux is entirely boundary layer has no slowly
expanding interior to supply fast wind.

\newcommand{\TabA}[1]{\parbox[t]{1.55in}{\raggedright #1\strut}}
\newcommand{\TabB}[1]{\parbox[t]{2.15in}{\raggedright #1\strut}}
\newcommand{\TabC}[1]{\parbox[t]{2.20in}{\raggedright #1\strut}}
\newcommand{\TabW}[1]{\parbox[t]{4.45in}{\raggedright #1\strut}}
\begin{deluxetable*}{lll}
\tablecaption{Axial-dipole (ADCH) and equatorial-dipole (EDCH) coronal holes
compared. Quantities marked $\dagger$ are computed here, each from a single
realization of the corresponding spectrum; those marked $\ddagger$ are measured
from sequences of observed maps (Table~\ref{tab:stability}); the remainder are
argued in Sections~\ref{sec:structure}--\ref{sec:wind}.\label{tab:summary}}
\tablehead{\colhead{Property} & \colhead{Axial-dipole hole} &
\colhead{Equatorial-dipole hole}}
\startdata
\multicolumn{3}{l}{\textbf{I. Congruent: no distinction}} \\
\hline
\TabA{Open area, pure dipole$^\dagger$} & \TabB{35.4\%} & \TabC{35.3\%} \\
\TabA{Core-tube $f_{ss}$, pure dipole$^\dagger$} & \TabB{1.724 (analytic 1.720)} & \TabC{1.720 (analytic 1.720)} \\
\TabA{Radial distribution of expansion$^\dagger$} & \multicolumn{2}{l}{\TabW{Completed fraction $[f(r)-1]/[f_{ss}-1]$ agrees to within $4\%$ above $1.4\,R_\odot$ across all four configurations}} \\
\TabA{Asymptotic wind at 1~au} & \multicolumn{2}{l}{\TabW{$\sim750\,\mathrm{km\,s^{-1}}$; indistinguishable}} \\
\hline
\multicolumn{3}{l}{\textbf{II. Divergent: genuine distinctions}} \\
\hline
\TabA{Relation to differential rotation} & \TabB{Boundary at constant latitude; photosphere slips along it} & \TabC{Boundary crosses latitudes; photosphere slips across its poleward flanks, held quasi-rigid by reconnection} \\
\TabA{Role in surface flux transport} & \TabB{Terminus; footpoint flux resides for years} & \TabC{Source region; footpoint flux dispersed by surface transport on the meridional crossing time} \\
\TabA{Stability of the dominant component$^\ddagger$} & \TabB{$g_{10}$ steady to $1\%$ over 18 days} & \TabC{$|g_{11}|$ varies by $21\%$ while forming, $8\%$ once established} \\
\TabA{Meridional flow at boundary} & \TabB{Convergent} & \TabC{Divergent} \\
\TabA{Filament channel} & \TabB{Polar crown; driven, persistent; hemispheric chirality} & \TabC{Sporadic; inherited, AR chirality; decays} \\
\TabA{Boundary topology} & \TabB{Single global separatrix} & \TabC{S-web QSLs, pseudostreamer separatrices} \\
\TabA{Rotational slip} & \TabB{$37^\circ$ and $67^\circ$/rot at $45^\circ$ and $60^\circ$; $30$--$73^\circ$ across the plausible band} & \TabC{$+5^\circ$/rot at the equator for the plotted rate, to $+13^\circ$ at the slow edge; sign not robust} \\
\TabA{Open area, realistic spectrum$^\dagger$} & \TabB{13.7\%} & \TabC{24.3\%} \\
\TabA{Deep interior$^\dagger$} & \multicolumn{2}{l}{\TabW{Reaches $28^\circ$ from the boundary at minimum against $24^\circ$ in the tilted-dipole case, with $23\%$ and $12\%$ of the open area in the interior. The two are close enough that we draw no conclusion from the comparison}} \\
\TabA{Non-monotonic tubes$^\dagger$} & \TabB{0.0\% of open area (this spectrum)} & \TabC{3.0\%; $N$ to 3.2, all within $2.3^\circ$ of boundary} \\
\TabA{Acceleration completed at $10\,R_\odot$} & \TabB{$\sim95\%$ assumed in Figure~\ref{fig:wind}; $64$--$80\%$ from near-Sun constraints at face value (Figure~\ref{fig:ips})} & \TabC{$\sim60\%$ measured; residual to $\sim70\,R_\odot$} \\
\hline
\multicolumn{3}{l}{\textbf{III. Expansion factor and its interpretation}} \\
\hline
\TabA{Median $f_{ss}$$^\dagger$} & \TabB{6.42} & \TabC{2.88} \\
\TabA{Core $f_{ss}$ (5th pct)$^\dagger$} & \TabB{4.36} & \TabC{1.75} \\
\TabA{Median $f_{\max}$$^\dagger$} & \TabB{6.43} & \TabC{2.90} \\
\TabA{Implication} & \multicolumn{2}{l}{\TabW{The equatorial-dipole hole expands less than the axial-dipole one, the opposite of what lateral confinement requires, and the completed fraction of the expansion is the same to within $4\%$; the geometry therefore offers no reason for its stream to reach only $60\%$ of its asymptotic speed at $10\,R_\odot$ \citep{Badman2026}}} \\
\TabA{Non-monotonic tubes as the acceleration source$^\dagger$} & \multicolumn{2}{l}{\TabW{Whether $3.0\%$ of the open area can carry a bulk stream is untested; the fraction itself depends on the assumed AR spectrum}} \\
\TabA{IPS phase-speed bias} & \multicolumn{2}{l}{\TabW{Factor 1.67 required, 2.78 available. All published apparent speeds lie inside the band of Figure~\ref{fig:wind}a, and \citet{Badman2026} independently favor this resolution, but the 1996 data are consistent with both readings; forward modeling remains the decisive test}} \\
\enddata
\tablecomments{The synthetic open-area and $\theta_b^{\max}$ entries are
properties of the spectra listed in Table~\ref{tab:metrics} and are computed at
a fixed truncation and seed radius; they are not directly comparable with
values derived from observed maps, for the reasons set out in
Section~\ref{sec:obsopen}.}
\end{deluxetable*}

\section{Conclusions} \label{sec:conclusions}

We have compared the two classes of coronal hole that a largely dipolar Sun can
produce: the axial-dipole coronal holes (ADCHs) of solar minimum, and the equatorial-dipole coronal holes (EDCHs) that form when the equatorial dipole leads the low-order field, as it did in 2025, when Parker Solar Probe found fast wind from such a hole still
accelerating at $10\,R_\odot$. We asked whether the coronal magnetic structure
of the two classes gives any reason to expect them to accelerate the wind
differently, using potential-field extrapolations of synthetic configurations
and of observed maps from the cycle 24/25 minimum, the 2024 maximum and the
2025 declining phase. Table~\ref{tab:summary} collects the comparison, grouped
by whether the configurations agree, differ, or yield a result that cuts
against the hypothesis under test. Our principal results are the following.
\begin{enumerate}
\item Pure axial and pure equatorial dipoles are exactly congruent under PFSS,
with open-area fractions of $35.4\%$ and $35.3\%$ and core expansion factors of
$1.724$ and $1.720$ against the analytic $1.720$; the small residuals are
numerical (Section~\ref{sec:congruence}). Any distinction between the
two coronal-hole classes must therefore arise from the orientation of the
boundary relative to the differential-rotation pattern, or from the multipolar
environment in which the dipole sits.
\item Persistence depends largely on orientation. Differential rotation leaves
an axial dipole intact but winds up and destroys an equatorial one, so the
axial-dipole hole is a stationary state at the terminus of surface flux
transport, where converging meridional flow sustains the polar crown filament
channel, while the equatorial-dipole hole depends on an equatorial dipole that
active-region emergence builds and flux transport dismantles.
\item The axial dipole is far steadier than the equatorial dipole, each
measured when it dominates the low-order field. Over $18$-day sequences of maps
the axial dipole at the cycle 24/25 minimum is steady to $1\%$, whereas the
equatorial dipole varies by $21\%$ near the 2024 maximum and $8\%$ in the 2025
declining phase, its orientation drifting far more slowly than the Sun rotates.
\item Not every low-latitude coronal hole is an equatorial-dipole hole. Near the
2024 maximum the field was quadrupole-led; only on the 2025 declining phase,
after the equatorial dipole doubled, did it approach a tilted dipole.
\item Flux-tube expansion does not predict slower acceleration from
equatorial-dipole holes. Their bulk expands \emph{less} than a polar hole at
every height, with a median $f_{ss}$ of $2.88$ against $6.42$, and the fraction of the expansion completed by any given height is the same in
every configuration to within $4\%$.
\item The one structural feature separating the classes is topological. The
equatorial-dipole boundary is pseudostreamer-dominated: $3\%$ of its open area
lies on flux tubes whose expansion is non-monotonic, peaking near
$1.27\,R_\odot$, and the axial-dipole configurations contain no such flux
tubes.
\item If the extended acceleration of equatorial-dipole wind is confirmed
against polar wind, it points to extended heating, such as the dissipation of reflection-driven Alfv\'en-wave turbulence over tens of solar radii, whose energy may be injected differently at the pseudostreamer-dominated boundary; both lie beyond a potential-field model. The scintillation evidence for rapid polar acceleration is itself open to a phase-speed bias.
\end{enumerate}

The lifetime of an equatorial-dipole hole is set by two components of the
field, which evolve on different timescales. The first is the strong field of
the active-region nests, which sustains the large-scale equatorial-dipole
component; it lasts as long as the nests do, and decays as their flux is
broken up and dispersed by convection. Because nests persist at fixed
longitudes while their flux emerges episodically, the configuration holds its
longitude while its strength follows each emergence and decay. The second is
the weak unipolar flux at the footpoints of the open field, in which the hole
itself is rooted; it persists until surface flux transport, through
differential rotation, meridional flow and supergranular diffusion, disperses
it or carries it into cancellation.

The expansion-factor comparison rests on the ordering of the expansion factors,
not their absolute values, which scale with the resolution of the photospheric
map. Within the equatorial-dipole configuration it is the small
multipole-associated holes around the main one that expand most
($f_{ss}=2.98$ against $2.64$), consistent with their role as the usual sources
of Alfv\'enic slow wind near maximum \citep{DAmicisBruno2015,WangKo2019}.
Whether the non-monotonic boundary population, only $3\%$ of the open area, can
carry a bulk stream is untested, and neither it nor the reconnection it implies
can be pursued within a potential-field model that forces the field radial at
$2.5\,R_\odot$.

Applied to observed photospheric field maps, the analysis also sets a limit on
what these extrapolations can decide. Open-area fraction, interior depth and
component count depend strongly on the harmonic truncation and on the height
at which field lines are released, and at matched truncation and height the
2024 and 2025 maps are not clearly distinguished by any of them, so we make no
claim that the open field reorganized between the two epochs. What did change,
measurably and without tracing, is the low-order field, a change the WSO series
confirms (Figure~\ref{fig:multipoles}) and \citet{Finley2026} trace to nests
emerging in 2025 January. The wind followed, from speeds barely above
$500\,\mathrm{km\,s^{-1}}$ to a stream at $750$--$800\,\mathrm{km\,s^{-1}}$, but
our extrapolations do not establish the chain connecting the two. The taxonomy
of Sections~\ref{sec:congruence} and~\ref{sec:structure} governs how a coronal
hole forms and how long it persists; what sets the speed of its wind is left
open to more detailed in situ analysis \citep[]{Rivera2026}.

The phase-speed bias remains adequate to reconcile the
interplanetary-scintillation constraint with the measured profile, requiring a
factor of $1.67$ where $2.78$ is available (Figure~\ref{fig:wind}a).
Establishing that it is the explanation, rather than merely a sufficient one,
will require comparing the predicted radial shape of the apparent speed against
published scintillation profiles under an explicitly stated tracer weighting.
Lateral confinement below the source surface is excluded as the source of the
extended acceleration; whether some equivalent process operates above it is a
question for global MHD modeling.

\begin{acknowledgments}
Wilcox Solar Observatory data used in this study were obtained via the web
site http://wso.stanford.edu, courtesy of J.~T.~Hoeksema. This work utilizes
data obtained by the Global Oscillation Network Group (GONG) program, managed
by the National Solar Observatory, which is operated by the Association of
Universities for Research in Astronomy (AURA), Inc.\ under a cooperative
agreement with the National Science Foundation. We acknowledge the use of
NASA/GSFC's Space Physics Data Facility's OMNIWeb service and OMNI data.

The SDO/HMI and SDO/AIA data are courtesy of NASA/SDO and the HMI and AIA
science teams; SDO is a mission for NASA's Living With a Star program. HMI
data were obtained through the Joint Science Operations Center (JSOC) at
Stanford University. 

Parker Solar Probe was designed, built, and is now operated by the Johns
Hopkins Applied Physics Laboratory as part of NASA's Living with a Star (LWS)
program (contract NNN06AA01C). Support from the LWS management and technical
team has played a critical role in the success of the Parker Solar Probe
mission. We acknowledge the SWEAP and FIELDS instrument teams, whose
measurements of the 2025 stream are used in Figures~\ref{fig:wind} and~\ref{fig:ips}.

This research was partially supported by the International Space Science Institute (ISSI) in Bern, through ISSI International Team project \#550 (Solar Sources and Evolution of the Alfv\'enic Slow Wind). AJF acknowledges support through the European Space Agency (ESA) Research Fellowship in Space Science. OAP acknowledges support from NSF SHINE 2229566 and NASA AIAH 80NSSC25K0386. STB and YJR were partially supported by Parker Solar Probe project through the SAO/SWEAP subcontract 975569.

\end{acknowledgments}

\appendix

\section{Harmonic Decomposition of the Photospheric Field}
\label{app:harmonics}

All harmonic coefficients quoted in this paper are obtained directly from the
photospheric radial field of each extrapolation, without tracing and without
regridding. Because the quantities in Tables~\ref{tab:stability}
and~\ref{tab:harmonic} carry much of the paper's observational weight, we set
out the procedure and the normalization convention explicitly.

\subsection{Grid and quadrature}

The SolarSoft extrapolations are evaluated on a grid of $N_\phi=384$ longitudes
and $N_\theta=192$ latitudes. The longitudes are uniform,
$\phi_i=2\pi i/N_\phi$. The latitudes are \emph{not} uniform: they are the
Gauss--Legendre nodes in $\mu\equiv\sin(\mathrm{latitude})=\cos\theta$. We
verify this for every file by comparing the stored latitude array against the
$192$-point Gauss--Legendre nodes, which agree to better than $10^{-4}$ degrees.

This makes the surface integral exact rather than approximate. For any function
$f$ on the sphere,
\begin{equation}
\oint f \,d\Omega \;=\; \int_0^{2\pi}\!\!\int_{-1}^{1} f\,d\mu\,d\phi
\;\simeq\; \frac{2\pi}{N_\phi}\sum_{i=1}^{N_\phi}\sum_{j=1}^{N_\theta}
w_j\, f(\mu_j,\phi_i),
\label{eq:quad}
\end{equation}
where $w_j$ are the Gauss--Legendre weights associated with the nodes $\mu_j$.
Gauss--Legendre quadrature with $N_\theta$ nodes integrates polynomials in
$\mu$ of degree up to $2N_\theta-1$ exactly, so for the degrees used here the
latitudinal integral carries no discretization error. The longitudinal sum is
exact for a periodic band-limited function. No $\cos(\mathrm{latitude})$ area
weighting is applied, because the $d\mu$ measure already accounts for it; using
$\sin\theta\,d\theta$ weighting on this grid would be incorrect.

\subsection{Basis functions}

We expand in real orthonormal spherical harmonics,
\begin{eqnarray}
Y_{lm} &=& \sqrt{2}\,N_{lm}\,P_l^{|m|}(\cos\theta)\,\cos(m\phi)
  \quad (m>0), \nonumber\\
Y_{l0} &=& N_{l0}\,P_l(\cos\theta), \label{eq:ylm}\\
Y_{lm} &=& \sqrt{2}\,N_{lm}\,P_l^{|m|}(\cos\theta)\,\sin(|m|\phi)
  \quad (m<0), \nonumber
\end{eqnarray}
with
\begin{equation}
N_{lm}=\sqrt{\frac{2l+1}{4\pi}\,\frac{(l-|m|)!}{(l+|m|)!}},
\label{eq:norm}
\end{equation}
and $P_l^m$ the associated Legendre functions without the Condon--Shortley
phase. These satisfy $\oint Y_{lm}Y_{l'm'}\,d\Omega=\delta_{ll'}\delta_{mm'}$;
evaluated with Equation~(\ref{eq:quad}) on the grid above we obtain
$\oint Y_{10}^2\,d\Omega=1.000000$ and
$\oint Y_{10}Y_{20}\,d\Omega\sim10^{-16}$, confirming both the basis and the
quadrature.

The coefficients follow by projection,
\begin{equation}
\tilde{g}_{lm}=\oint B_r(R_\odot,\theta,\phi)\,Y_{lm}(\theta,\phi)\,d\Omega .
\label{eq:proj}
\end{equation}

\subsection{Normalization}

Equation~(\ref{eq:proj}) returns coefficients in the orthonormal convention,
which is not the convention usual in solar physics. For a pure axial dipole
$B_r=B_0\cos\theta$ of polar field strength $B_0$, Equation~(\ref{eq:proj})
gives $\tilde{g}_{10}=B_0\sqrt{4\pi/3}=2.047\,B_0$ rather than $B_0$. We
therefore convert to Schmidt semi-normalized coefficients,
\begin{equation}
g_{lm}=\tilde{g}_{lm}\Big/\sqrt{\frac{4\pi}{2l+1}},
\label{eq:schmidt}
\end{equation}
so that $g_{10}$ is the polar field strength of the equivalent axial dipole and
all coefficients are in gauss. Every value quoted in this paper uses
Equation~(\ref{eq:schmidt}). This matters for comparison with published series:
values in the orthonormal convention exceed Schmidt values by $2.047$ at $l=1$
and $1.585$ at $l=2$.

\subsection{Derived quantities}

For $m\ne0$ the cosine and sine coefficients form a vector in the equatorial
plane, and we quote its magnitude and orientation,
\begin{equation}
|g_{lm}|=\sqrt{g_{l,m}^2+g_{l,-m}^2},
\qquad
\varphi_{lm}=\frac{1}{m}\arctan\!\left(\frac{g_{l,-m}}{g_{l,m}}\right).
\label{eq:vector}
\end{equation}
For $l=m=1$ these are the strength and the Carrington longitude of the
equatorial dipole; $\varphi_{11}$ is the phase whose drift is reported in
Table~\ref{tab:stability}. Power fractions are
\begin{equation}
\Pi_l=\frac{\sum_{m=-l}^{l}g_{lm}^2}{\sum_{l'}\sum_{m=-l'}^{l'}g_{l'm}^2},
\label{eq:power}
\end{equation}
with the denominator taken over $l'\le3$ where stated and over $l'\le20$
otherwise; the two differ, and we give the range used with each quoted value.
Because $\Pi_l$ is a ratio of coefficients at the same $l$ in the numerator, it
is unaffected by the choice between Equations~(\ref{eq:proj})
and~(\ref{eq:schmidt}) only when the denominator is restricted to a single $l$;
for the multi-$l$ denominators used here the Schmidt values are used
throughout.

\subsection{Sensitivity}

Three points bear on how these numbers should be read.

The decomposition uses only $B_r$ at $r=R_\odot$ and is therefore independent
of the source-surface radius, the field-line tracing, and the seed height, all
of which affect the open-field quantities discussed in
Section~\ref{sec:obsopen} but none of which enter here.

The far-side field of our maps is modelled rather than observed
(Section~\ref{sec:observed}), and it contributes to the low-order coefficients.
The five-map sequences of Table~\ref{tab:stability} bound the resulting
variability empirically: at solar minimum, when little flux is present on
either hemisphere, $g_{10}$ is stable to $1\%$ over eighteen days, which shows
that the larger scatter at maximum is not a generic artifact of the map
product.

Coefficients at $l\ge2$ with $m\ne0$ are the least stable in all three
intervals, $|g_{22}|$ varying by $13$--$35\%$ depending on phase, and we base
no argument on that coefficient.

\end{document}